\documentclass{article}

\usepackage{arxiv}

\usepackage[utf8]{inputenc}
\usepackage[T1]{fontenc}   
\usepackage{hyperref}      
\usepackage{url}           
\usepackage{booktabs}      
\usepackage{amsfonts}      
\usepackage{amssymb}
\usepackage{bm}
\usepackage{amsmath}
\usepackage{textcomp, gensymb}
\usepackage{array,multirow}
\usepackage{nicefrac}      
\usepackage{microtype}     
\usepackage{lipsum}		
\usepackage{graphicx}
\usepackage{float}
\usepackage[numbers]{natbib}
\usepackage{doi}
\usepackage{placeins}

\title{Implicit neural representations for unsupervised super-resolution and denoising of 4D flow MRI}

\author{\href{https://orcid.org/0000-0002-3974-5945}{Simone Saitta}\thanks{Correspondence: simone.saitta@polimi.it} \\
	Department of Electronics, Information and Bioengineering\\
	Politecnico di Milano\\
	Milan, Italy \\
	\texttt{simone.saitta@polimi.it}
        \And
        Marcello Carioni \\
        Department of Applied Mathematics\\ University of Twente \\
        7500AE Enschede, The Netherlands \\
        \texttt{m.c.carioni@utwente.nl} \\
        \And
        Subhadip Mukherjee \\
        Department of Computer Science \\
        University of Bath \\
        Bath, UK \\
        \And
        Carola-Bibiane Sch{\"o}nlieb \\
        Department of Applied Mathematics and Theoretical Physics \\ University of Cambridge \\
        Cambridge, UK \\
        \And
        Alberto Redaelli \\
        Department of Electronics, Information and Bioengineering\\
	Politecnico di Milano\\
	Milan, Italy \\
 }

\renewcommand{\shorttitle}{ }

\begin{document}
\maketitle

\begin{abstract}
 \textit{Context}. 4D flow magnetic resonance imaging (4D flow MRI) is the only non-invasive imaging method that can provide time-resolved measurements of blood flow velocities. However, velocity fields detected by this technique have significant limitations that prevent accurate quantification of blood flow markers. These limitations are mainly related to low spatio-temporal resolution and measurement noise. Several lines of research have been pursued to overcome the main limitations. Among these, coordinate-based neural networks have shown to be able to represent complex signals as continuous functions, making them suitable for super-resolution tasks. In this work we investigate sinusoidal representation networks (\textsc{siren}s) for time-varying 3-directional velocity fields measured in the aorta by 4D flow MRI, achieving denoising and super-resolution.\\
 \textit{Method}. We trained our method on 4D voxel coordinates and enforce the no-slip condition at the vessel wall. First, we benchmarked our approach using synthetic measurements generated from a computational fluid dynamics (CFD) simulation for which ground truth velocity fields are available. Three different levels of noise were simulated: \textit{mild}, \textit{medium} and \textit{extreme}. Then, we test our method on a real 4D flow MRI scan of a patient with aneurysm of the ascending aorta. We assessed both velocity and wall shear stress (WSS) fields obtained with our method. \\
 \textit{Results}. The performance of different \textsc{siren} architectures was evaluated on synthetic measurements. The best configuration was chosen as the one that minimized the sum of velocity normalized root mean square error (vNRMSE), magnitude normalized root mean square error (mNRSME) and direction error (DE). A \textsc{siren} with 20 layers and 300 neurons per layer gave the lowest error for all levels of noise. The results of the \textsc{siren} with the chosen configuration were compared against state-of-the-art methods for 4D flow denoising and super-resolution. Our approach outperformed all existing techniques, giving up to 50\% lower vNRMSE, 42\% lower mNRMSE and 15\% lower DE on velocity and WSS fields. Additionally, the developed approach produced denoised and super-resolved velocity fields from clinical data, while maintaining low discrepancies in macroscopic flow measurements.\\
 \textit{Conclusions}. We showed the feasibility of \textsc{siren}s to representing complex and multi-dimensional blood flow velocity fields obtained from 4D flow MRI. Our approach is both quick to execute and straightforward to implement for novel cases. By meticulously optimizing our \textsc{siren} architecture, we leverage the spectral bias to generate a functional representation of our data with minimized noise, surpassing current solutions. Our method produces continuous velocity fields that can be queried at any spatio-temporal location, effectively achieving 4D super-resolution.
\end{abstract}


\section{Introduction}\label{sec:introduction}
Accurate hemodynamic assessment is essential for having a deeper understanding of cardiovascular pathophysiology \citep{davies2009hemodynamic, bagan2006cerebral}. For certain cardiovascular conditions such as aortic coarctation, valvular disfunction and vascular aneurysm, disease diagnosis and management are based on \textit{in vivo} hemodynamic biomarkers \citep{writing20222022, bertoglio2018relative} that can be obtained by either catheter insertion or non-invasive blood flow imaging methodologies.
To date, 4-dimensional flow encoded magnetic resonance imaging (4D flow MRI or 4D flow) is the only existing non-invasive imaging technique that provides true time-resolved 3-dimensional (3D) and 3-directional blood flow velocity measurements \citep{markl20124d}. 4D flow is based on the phase contrast MRI (PC-MRI) principle, which makes use of bipolar magnetic gradients to calculate the phase shift of moving protons. PC-MRI encodes tissue velocity $\mathbf{v}(\mathbf{x}, t) \in \mathbb{R}^3$ at spatial location $\mathbf{x}$ during cardiac phase $t (1 \leq t \leq N_t)$ according to:
\begin{equation}\label{eq:complex_img}
    \rho_i(\mathbf{x},t) = \rho_0(\mathbf{x}, t)\exp \left( j\pi\frac{(\mathbf{\Phi}\mathbf{v}(\mathbf{x}, t)))_i}{VENC} \right),
\end{equation}
where $VENC$ is a manually set parameter determining the maximum velocity that can be recorded, $i=0, ..., 3$ are the encoded velocity components, and adopting a four-point velocity encoding, $\mathbf{\Phi}$ is defined as:
\begin{equation}
 \mathbf{\Phi} = 
    \begin{bmatrix}
    0 & 0 & 0 \\
    1 & 0 & 0 \\
    0 & 1 & 0 \\
    0 & 0 & 1 
    \end{bmatrix}
    .
\end{equation}
Hence, the measured tissue velocity component $i$ is proportional to the phase shift of the reconstructed images $\rho_i$ \citep{vishnevskiy2020deep}.\\
Considering $\bm{\rho}_{it} \in \mathbb{R}^{N_r \times N_c \times N_s}$ a discretized complex PC image on a Cartesian $N_r \times N_c \times N_s$ grid corresponding to cardiac phase $t$ and velocity component $i$, the reconstructed image $\mathbf{H}(\bm{\rho}_{it}) \in \mathbb{R}^{N_r \times N_c \times N_s}$ can be modeled as:
\begin{equation}
   \mathbf{H}(\bm{\rho}_{it}) = \mathbf{F}^{-1}(\mathbf{M}(\mathbf{F}(\bm{\rho}_{it}) + \epsilon)),
\end{equation}
where $\mathbf{F}$ is the Fourier transform, $\mathbf{M} \in \{0, 1\}^{N_r \times N_c \times N_s}$ defines the undersampling mask in \textit{k}-space, and $\epsilon \in \mathbb{C}^{N_r \times N_c \times N_s}$ is the additive complex noise. Herein, we have neglected coil sensitivity maps for simplicity, a more rigorous description of the PC-MRI measurement operator can be found in \citep{vishnevskiy2020deep}.\\
Velocity fields measured by 4D flow MRI can be processed after reconstruction to quantify more complex and clinically relevant hemodynamic biomarkers such as wall shear stress (WSS) \citep{barker2012bicuspid, trenti2022wall, bissell2013aortic}, relative pressure \citep{bock2011vivo, saitta2019evaluation} and vortex structure \citep{reiter2015blood}.
Nonetheless, the large amount of raw data collected during a 4D flow MRI scan limits this imaging technique, entailing longer image reconstruction time and more difficult image analysis with respect to standard magnetic resonance angiography (MRA). Modern compressed sensing has enabled substantial shortening of 4D flow MRI acquisition time \citep{bollache2018k, rich2019bayesian} by undersampling \textit{k}-space data and exploiting priors about data regularity during reconstruction \citep{kim2012accelerated, valvano2017accelerating}. Although modern techniques could achieve 4D flow MRI reconstructions in under 5 minutes \citep{ma2019aortic}, the resulting flow images still suffer from important limitations that make them inadequate for accurate quantification of blood flow markers. These limitations mainly concern spatio-temporal resolution, velocity encoding (VENC) related signal to noise ratio (SNR) and \textit{k}-space noise \citep{jiang2015quantifying}. For cardiothoracic acquisitions, image spacing is usually isotropic and in the range of 1.5-3 mm\textsuperscript{3}, whilst typical temporal resolution is 30-50 ms \citep{dyverfeldt20154d}. Furthermore, 4D flow MR images are corrupted by noise, which is commonly assumed to be zero-mean Gaussian in frequency space \citep{cardenas2008noise}. 
Overcoming 4D flow limitations by reducing noise levels and enhancing its spatio-temporal resolution, would lead to more accurate hemodynamic assessment, boosting its widespread adoption and increasing its clinical usefulness. 
In this work, we present a novel application of implicit neural representations (INRs) \citep{sitzmann2020implicit} to achieve super-resolution (SR) and denoising of 4D flow MRI velocity fields. Our approach provides a continuous reconstruction of blood flow in both space and time, showing superior performance with respect to state-of-the-art methods for 4D flow denoising. 

\subsection{Related Work}
Over the past two decades, several lines of research have been pursued to overcome the main limitations of 4D flow MRI velocity measurements and provide more accurate non-invasive hemodynamic assessment. Herein, we categorize these efforts broadly into two groups: model-based and data-based approaches. Model-based methods are based on computational fluid dynamics (CFD) informed using MR images, whereas data-based methods directly operate on image data and apply regularized interpolation to enhance flow-encoded images \citep{funke2019variational, koltukluouglu2018boundary, kontogiannis2022joint, kontogiannis2022physics, ong2015robust, busch2013reconstruction, fathi2020super}.

\subsubsection{Model-based approaches}
As an alternative to 4D flow MRI, CFD has been widely applied to study cardiovascular flows \citep{marsden2009evaluation, nannini2021aortic, morris2016computational, pirola20194}. In contrast to 4D flow, CFD simulations can provide noise-free blood flow velocity fields at arbitrarily high spatio-temporal resolutions. By solving the Navier-Stokes (N-S) equations, numerical hemodynamic solutions inherently satisfy mass and linear momentum conservation laws. Nevertheless, when modeling subject specific cardiovascular flows, the accuracy of CFD simulations heavily relies on the choice of boundary conditions \citep{armour2021influence, pirola2017choice} and blood constitutive model \citep{xiang2012newtonian, doost2016numerical}. When comparing CFD solutions to 4D flow velocity measurements, these assumptions inevitably lead to substantial discrepancies between hemodynamic markers computed via the two approaches \citep{naito2012magnetic, jain2016transitional, jiang2011flow}.
Model-based approaches typically formulate their approach as an inverse N-S problem, in which one or more unknown parameters in the governing partial differential equations (PDEs) are optimized to minimize an objective functional representing the discrepancy between the N-S solution and the measured data. Model-based methods effectively address the limitations of noise and low spatio-temporal resolution of flow measurements, yielding velocity fields defined on fine computational grids and at arbitrarily small time steps (\textless 10 ms).\\
In the seminal works of Funke et al. \citep{funke2019variational} and Koltukluo\u{g}lu et al. \citep{koltukluouglu2018boundary}, this approach came down to solving a N-S boundary value problem formulated in a variational data assimilation framework, in which one or more boundary conditions are optimized in either 3D \citep{koltukluouglu2018boundary} or 4D \citep{funke2019variational}.  
Recently, a more comprehensive approach was proposed by Kontogiannis et al. \citep{kontogiannis2022joint} to jointly recover both boundary conditions, domain boundary and kinematic viscosity by assimilating noisy flow image data into the N-S solution. The same well-designed settings were later extended by the same authors to cope with undersampled synthetic phase contrast data \citep{kontogiannis2022physics}, resulting in the first N-S informed compressed sensing reconstruction method. However, their method was only applied to 2D cases with steady flow conditions. In fact, despite the considerable progress made, all the mentioned techniques are computationally intensive, and their cost can enormously increase in higher dimensions, undermining their feasibility for real medical case-scenarios. For this reason, there is still lack of model-based approaches that can efficiently assimilate time-dependent 3D velocity fields from real medical flow images for concrete human data applications.

\subsubsection{Data-based approaches}
In contrast to model-based approaches, data-based methods do not rely on solving the governing PDEs. Instead, they employ interpolation techniques to approximate flow data, incorporating some form of regularization to impart desirable characteristics to the resulting velocity fields. Within this category, we discern between conventional denoising approaches and recent neural network-based methods. 

\paragraph{Conventional denoising approaches} adopt signal processing tools to enhance the acquired flow MRI data after reconstruction. To incorporate prior physical knowledge in their processing stage, several works have exploited the assumption of blood incompressibility. This physical condition is usually enforced through a divergence-free constraint on the reconstructed velocity field \citep{busch2013reconstruction, ong2015robust}. Busch et al. \citep{busch2013reconstruction}, achieved effective denoising by projecting noisy flow measurements onto a 3D space of divergence-free radial basis functions (RBFs). Moreover, their approach allows for incorporation of boundary conditions in the reconstructed flow field. Nonetheless, strict enforcement of the divergence-free condition may result in over-regularized velocity fields, especially in flow regions near edges of flow (boundaries). 
An approach to \textit{softly} enforce the divergence-free condition was proposed by a Ong et al. \citep{ong2015robust}. By constructing divergence-free wavelets (DFWs), the authors were able to decompose measured velocity fields into divergence-free and non-divergence-free wavelet coefficients, promoting the former and penalizing the latter according to suitably tuned thresholds. The described process is essentially analogous to wavelet denoising, but the different choice of wavelets enables correction of divergent flow components, providing flow fields with more coherent streamlines when compared to unfiltered medical data. 
By directly processing images, data-based methods are in general more computationally efficient than model-based techniques. However, none of the proposed approaches so far has taken into account the time-dependent nature of 4D flow data. Additionally, working on structured grids (images) prevents a precise quantification of near-wall blood flow markers that are often sought-after in medical applications, such as WSS fields.
In a recent study \citep{thewlis20234}, 4D RBFs were used to reconstruct velocity fields measured with MRI. To make their approach computationally light, the authors combined multi-quadric RBF interpolation with a partition of unity scheme. Since it operates on scattered data, this approach results efficient in high dimensions, and well-suited for enforcing Dirichlet boundary conditions on the reconstructed velocity fields and improving quantification of near-wall features.  

\paragraph{Neural network approaches} are a relatively new class of data-based methods that can be thought of as approximations of complex nonlinear functions. Within the scientific computing community, a popular paradigm for exploiting the high expressive power of neural networks (NNs) is represented by physics informed neural networks (PINNs). In the formulation proposed in \citep{raissi2019physics}, PINNs are parametrized multilayer perceptrons (MLPs) that learn a function mapping coordinates to outputs. PINNs can be seen as continuous functions fitting sparse observations and simultaneously minimizing the residual of a PDE (in differential form) that is identified as the mathematical model that generated the data. This is done by constructing the loss function as a weighted sum of two terms: a data fidelity term and a PDE residual. By leveraging automatic differentiation, evaluation of the PDE residual does not introduce severe numerical errors, albeit entailing considerable increase in computational cost. 
The relative ease of implementation of PINNs and their potential to seamlessly combine measurements with PDEs has made this approach appealing for incorporating the N-S equations in fluid flow reconstruction applications \citep{jin2021nsfnets, raissi2020hidden, jin2020time, wang2017physics}. Nonetheless, when training PINNs, choosing the correct hyperparameters is often the result of a trial and error procedure. When dealing with high dimensional problems, hyperparameter tuning can become unfeasible. In particular: \textit{i}) a sufficient number of collocation points (coordinates where the PDE residual is evaluated) needs to be considered to achieve good accuracy \citep{raissi2019physics}, and \textit{ii}) the relative weights of the PDE residual in the loss function needs to be carefully calibrated \citep{wang2021understanding}. These issues make PINNs still difficult to apply to real-world fluid mechanics applications where data is often noisy and high dimensional.   
To date, the only application of PINNs to 4D flow MRI data is the work of Fathi et al. \citep{fathi2020super}. The authors tested their approach on synthetic 4D flow data from a reference CFD simulation for which ground truth velocity and pressure fields were known, achieving SR and superior denoising with respect to DFWs \citep{ong2015robust}. Successively, they evaluated their method on \textit{in vitro} 4D flow velocity measurements, but not on \textit{in vivo} medical data. \\
PINNs can be thought of as a subset of coordinate-based MLPs \citep{sitzmann2020implicit, tancik2021learned, martel2021acorn}. These networks can represent complex signals by taking low-dimensional coordinates as input and returning the value of the signal at the input locations. Such signal representations are often referred to as INRs \citep{sitzmann2020implicit}. Signals represented by coordinate-based MLPs can be orders of magnitude more compact than their grid-based counterparts \citep{tancik2020fourier}. Coordinate-based MLPs are known to suffer from severe spectral bias, namely they struggle to learn the high frequency signal components. To overcome this limitation, recent studies have proposed the introduction of a sinusoidal mapping of input coordinates \citep{tancik2020fourier, zhong2019reconstructing, mildenhall2021nerf}. In practice, this consists in a Fourier feature encoding \citep{rahimi2007random} of input coordinates $\mathbf{x}$ to $\gamma(\mathbf{x}) = [a_1\cos(2\pi\mathbf{b}_1^\intercal \mathbf{v}), a_1\sin(2\pi\mathbf{b}_1^\intercal \mathbf{v}), ..., a_m\cos(2\pi\mathbf{b}_m^\intercal \mathbf{v}), a_m\sin(2\pi\mathbf{b}_m^\intercal \mathbf{v})]$, where, in most cases, $a_j =1$ and $\mathbf{b}_j$ is sampled from an isotropic distribution \citep{tancik2020fourier}. Successful biomedical applications of coordinate-based networks include 3D vascular surface reconstruction \citep{alblas2023going} and non-rigid medical image registration \citep{wolterink2022implicit}. A further improvement of coordinate-based MLPs to capture high order derivatives in the output signal was achieved by employing sinusoidal activation functions for every hidden neural network layer, as introduced by Sitzmann et al. \citep{sitzmann2020implicit}. Sinusoidal representation networks (\textsc{siren}s) have been shown to be suited for representing complex natural signals, including images, solutions to Poisson equations and 3D shapes \citep{sitzmann2020implicit}.

In this work, we employ \textsc{siren}s for learning time-varying velocity fields measured by 4D flow MRI. We leverage the MLP's architecture to introduce an implicit prior to constrain the space of solutions and investigate such implicit regularization bias towards lower frequencies, which simultaneously prevents overfitting and reduces noise in flow-encoded MR images.

\section{Method}

\subsection{Problem setting}\label{sec:setting}
Partially borrowing the notation from \citep{kontogiannis2022physics}, from here on we will use the superscripts $(\cdot)^*$ to denote measured quantities and $(\cdot)^\bullet$ to denote ground truth quantities (when they exist). Additionally, quantities defined on unstructured point sets (or meshes) will be denoted with lowercase letters and parentheses, while quantities defined on structured grids (images) will be denoted with uppercase letter and square brackets.\\
Let $V^* \in \mathbb{R}^{N_r \times N_c \times N_s \times N_t \times 3}$ be a time-resolved flow image volume sequence obtained from a reconstructed 4D flow acquisition, with $N_r$, $N_c$, $N_s$ and $N_t$ being the number of rows, columns, slices and cardiac frames, respectively. Let $\Delta \bm{x} \in \mathbb{R}^3$ and $\Delta t \in \mathbb{R}$ be the physical spacings between adjacent voxels along the spatial and temporal dimensions. At a voxel center with 4D coordinates $[\bm{x}_i, t_j]$, where $\bm{x}_i \in \mathbb{R}^3$, we denote the measured velocity vector as: $V^*[\bm{x}_i, t_j] \in \mathbb{R}^3$, with $i \in [1, N_r \times N_c \times N_s]$, and $j \in [1, N_t]$. Herein, the index $i$ refers to the flattened spatial coordinates of the image.\\
We are interested in representing $V^*$ with a continuous function $f : \mathbb{R}^4 \to \mathbb{R}^3$, where $f(\bm{x}_i, t_j) = V^*[\bm{x}_i, t_j]$. To approximate $f$, we use an MLP $f_\mathbf{\Theta}$ with weights $\mathbf{\Theta}$ and with sinusoidal activation functions (\textsc{siren}). The $l$-th layer of a \textsc{siren} receiving a generic input tensor $\mathbf{x}_l \in \mathbb{R}^{Q_l}$ performs the following operation:
\begin{equation}
    \mathbf{x}_{l+1} = \sin(\mathbf{\Theta}_l \mathbf{x}_l + \mathbf{b}_l),
\end{equation}
where $\mathbf{\Theta}_l \in \mathbb{R}^{P_l \times Q_l}$ and $\mathbf{b}_l \in \mathbb{R}^{P_l}$ are the weight matrix and biases of the $l$-th layer, respectively. Following \citep{sitzmann2020implicit}, each weight $\theta$ is initialized so that $\theta \sim \mathcal{U}(-\sqrt{6/c}, \sqrt{6/c})$, where $c$ is the generic input feature size. Furthermore, as proposed by \citep{sitzmann2020implicit}, the first layer of the \textsc{siren} is modified as: $\sin(\omega_0 \cdot \mathbf{\Theta}\mathbf{x}+\mathbf{b})$. Following \citep{sitzmann2020implicit}, we set $\omega_0=30$.\\
For our purposes, a key advantage of this formulation lies in representing a high dimensional image as a continuous function that can be queried at arbitrary spatio-temporal resolutions.

\subsection{Training a SIREN in 4D}\label{sec:training}
In most practical cases dealing with blood vessels, one is only interested in reconstructing blood flow within a bounded region $\Omega \subset \mathbb{R}^3$ with inflow boundary $\Gamma_{\mathrm{i}}$, outflow boundary $\Gamma_{\mathrm{o}}$ and wall boundary $\Gamma_{\mathrm{w}}$, and within a time interval $[t_a, t_b]$. The inner fluid region is denoted by $\Omega_\mathrm{f} = \Omega \setminus \{ \Gamma_\mathrm{i} \cup \Gamma_\mathrm{o} \cup \Gamma_\mathrm{w} \}$. In our approach, velocity field reconstruction is achieved by sampling $N_{\mathrm{f}}$ spatial voxel coordinates $\bm{x}^{(\mathrm{f})}$ from $\Omega_{\mathrm{f}}$ repeated over the time interval $[t_a, t_b]$, i.e., $(\bm{x}_i^{(\mathrm{f})}, t_j)$, with $i=1,2,...,N_{\mathrm{f}}; j=1,2,...,N_t$. Additionally, we enforce the no-slip condition on the vessel wall by sampling $N_{\mathrm{w}}$ spatial coordinates from $\Gamma_\mathrm{w}$ repeated over the time interval $[t_a, t_b]$, i.e., $(\bm{x}_p^{(\mathrm{w})}, t_q)$, with $p=1,2,...,N_{\mathrm{w}}; q=1,2,...,N_t$.
Of note, we oversample spatial coordinates from $\Gamma_{\mathrm{w}}$ by setting $N_{\mathrm{f}} \approx N_{\mathrm{w}}$. We denote the total number of spatial coordinates used for \textsc{siren} training as $N = N_{\mathrm{f}} + N_{\mathrm{w}}$.
Unlike the original \textsc{siren} formulation, this approach allows us to only use a relatively small set of coordinates compared to the total number of image points, greatly reducing training time and memory cost.

For each generic input coordinate pair $(\bm{x}, t)$, the following non-dimensionalization is performed:
\begin{equation}\label{eq:adim}
    \hat{\bm{x}} = \frac{\bm{x} - \bm{x}_{min}}{ \Delta \bm{x}} D, \quad \hat{t} = \frac{t - t_{min}}{ \Delta t} D,
\end{equation}
where $\bm{x}_{min}$ and $t_{min}$ are the minimum spatial and temporal coordinates, and we set $D=0.01$.\\
Training $f_\mathbf{\Theta}$ implies solving the following minimization problem: 
\begin{equation}\label{eq:train}
   \min_{\mathbf{\Theta}} \mathcal{L}(\mathbf{\Theta}),
\end{equation}
where the loss function $\mathcal{L}$ is given by the misfit between the MLP prediction and the measured data plus a boundary condition term:
\begin{equation}
    \mathcal{L}(\mathbf{\Theta}) = \sum\limits_{i=1, j=1}^{N_{\mathrm{f}}, N_t} \lVert f_\mathbf{\Theta}(\bm{\hat{x}}^{(\mathrm{f})}_i, \hat{t}_j) - V^*[\bm{x}^{(\mathrm{f})}_i, t_j] \rVert_2^2 + 
    \sum\limits_{p=1, q=1}^{N_{\mathrm{w}}, N_t} \lVert f_\mathbf{\Theta}(\bm{\hat{x}}^{(\mathrm{w})}_p, \hat{t}_q)\rVert_2^2.
\end{equation}
Hence, the only supervision comes from the image values and fixed Dirichlet boundary conditions, making the approach fully unsupervised. 
Training is carried out using a limited memory Broyden-Fletcher-Goldbarb-Shanno (L-BFGS) algorithm with learning rate of 1 until $\nabla_\mathbf{\Theta} {\mathcal{L}} = 0$.

\subsection{SIREN evaluation}
Once trained, $f_\mathbf{\Theta}$ can be queried at arbitrary spatio-temporal collocation points denoted as $(\bm{x}', t')$, sampled from the spatio-temporal domain $\Omega \times [t_a, t_b]$. To evaluate the generalization capabilities of a trained \textsc{siren}, collocation points corresponding to $N'$ spatial coordinates and $N'_t$ temporal coordinates, i.e., $(\bm{x}'_n, t'_n)$, with $n=1,2,...,N'; m=1,2,...,M'$, are defined at $\approx \times$20 higher spatial resolution and $\times$10 higher temporal resolution than image coordinates. Non-dimensionalization of evaluation coordinates is operated consistently with the one used for the training set:
\begin{equation}
        \hat{\bm{x}}' = \frac{\bm{x}' - \bm{x}_{min}}{ \Delta \bm{x}} D, \quad \hat{t}' = \frac{t' - t_{min}}{ \Delta t} D.
\end{equation}
Therefore, we evaluate $f_\mathbf{\Theta}$ on: $(\hat{\bm{x}}'_n, \hat{t}'_m), n=1,...,N'; m=1,...,M'$.

\subsection{Error quantification}
To quantify errors obtained in experiments, three different metrics were used. Differences between a reference vector field $\bm{u}_{ref}$ and another generic vector field $\bm{u}$, were evaluated by computing magnitude and vector normalized-root-mean-squared-errors (mNRMSE and vNRMSE, respectively), and the direction error (DE) as:
\begin{equation}\label{eq:sNRSME}
    mNRMSE = \frac{1}{\max{\lvert \bm{u}_{ref}}\rvert} \sqrt{\frac{1}{K} \sum \limits_{k=1}^{K} (\lvert \bm{u} \rvert - \lvert \bm{u}_{ref} \rvert)^2_k},
\end{equation}

\begin{equation}\label{eq:vNRSME}
    vNRMSE = \frac{1}{\max{\lvert \bm{u}_{ref}}\rvert} \sqrt{\frac{1}{K} \sum \limits_{k=1}^{K} (\bm{u} - \bm{u}_{ref})^2_k},
\end{equation}

\begin{equation}\label{eq:DE}
    DE = \frac{1}{K} \sum \limits_{k=1}^{K} \left( 1 - \frac{\lvert \bm{u}_{ref,k} \cdot \bm{u}_{k} \rvert}{\lvert \bm{u}_{ref,k} \rvert  \lvert \bm{u}_{k} \rvert} \right), 
\end{equation}
where $K$ is the generic number of 4D points where the two velocity fields are evaluated. 

\subsection{Wall shear stress analysis}\label{sec:wss_method}
From a generic velocity field, the WSS field was calculated following the approach described in \citep{petersson2012assessment}. From the definition of WSS for a Newtonian fluid:
\begin{equation}\label{eq:wss}
    WSS = \mu \left( \frac{\partial{v}}{\partial{y}} \right)_{y=0},
\end{equation}
where $\mu$ is the dynamic viscosity, $v$ is the component of the velocity vector that is locally parallel to the wall, and $y$ is the Euclidean distance from the wall, for each spatial 3D point on the vessel wall ($\Gamma_\mathrm{w}$), the implemented WSS calculation method requires interpolation of the velocity fields at 2 points evenly spaced by a distance $\delta_n$ along the inward normal. The so obtained local velocity profile is interpolated with a quadratic function, whose analytical derivative is used to approximate $\partial{v}/\partial{y}$. In our experiments, we set $\delta_n=0.5 mm$

\subsection{Case 1: synthetic 4D flow MRI}\label{sec:synth_method}
\begin{figure}[htbp]
    \centering
    \includegraphics[width=\textwidth]{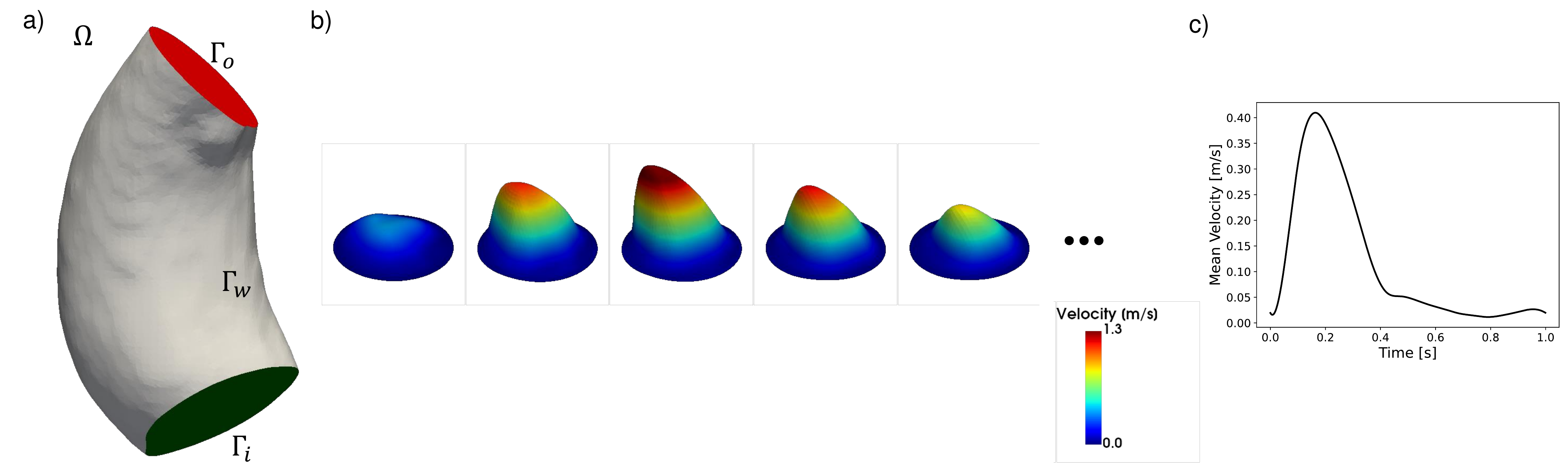}
    \caption{a) Computational domain. b) Time-varying 3-directional velocity profiles prescribed at as inlet boundary condition. c) Flow waveform imposed at $\Gamma_\mathrm{i}$} 
    \label{fig:cfd_setup}
    \end{figure}

    \paragraph{CFD simulation.}
    To have a benchmark for evaluating the proposed method, a synthetic 4D flow MRI acquisition was created from a reference CFD simulation. First, the ascending aorta of a subject with thoracic aorta aneurysm (TAA) was segmented from 3D MRA images using open-source software \citep{yushkevich2016itk}. The segmented domain $\Omega$, was divided into 3 subdomains: inlet $\Gamma_\mathrm{i}$, outlet $\Gamma_\mathrm{o}$ and wall $\Gamma_\mathrm{w}$ (Figure \ref{fig:cfd_setup}a). A 3D tetrahedral mesh with a base size of 0.6 mm was generated using the \textit{vmtk} library \citep{izzo2018vascular}. The final volumetric mesh consisted of $\approx$ 800k nodes. Time-varying 3-directional velocity profiles (Figure \ref{fig:cfd_setup}b) were prescribed as inlet boundary conditions, mapping a realistic TAA inlet velocity profile to the $\Gamma_\mathrm{i}$ following the approach described in \citep{saitta2022data} and producing the flow waveform represented in Figure \ref{fig:cfd_setup}c. A zero-pressure condition was enforced on $\Gamma_\mathrm{o}$ and a homogeneous Dirichlet boundary condition (no-slip) was assumed on $\Gamma_\mathrm{w}$. Blood was modeled as a Newtonian fluid with constant density $\rho$ = 1060 kg/m\textsuperscript{3} and dynamic viscosity $\mu$ = 0.004 Pa$\cdot$s.
    A finite volume simulation was run at a fixed time step of 0.001 s. An implicit scheme with splitting of operators (PISO) was used to solve the governing equations of blood flow in Star-CCM+. Results were exported at every timestep within the time interval [0.2 - 0.399] s (peak to late systole), yielding a sequence of $M'=200$ velocity fields $\bm{u}^\bullet$ defined on the computational nodes $\bm{x}'_n, n=[1, ..., N']$ over the simulation time coordinates $t'_m, m=[1, ..., M']$.
    
    \FloatBarrier

\begin{figure}[htbp]
    \centering
    \includegraphics[width=\textwidth]{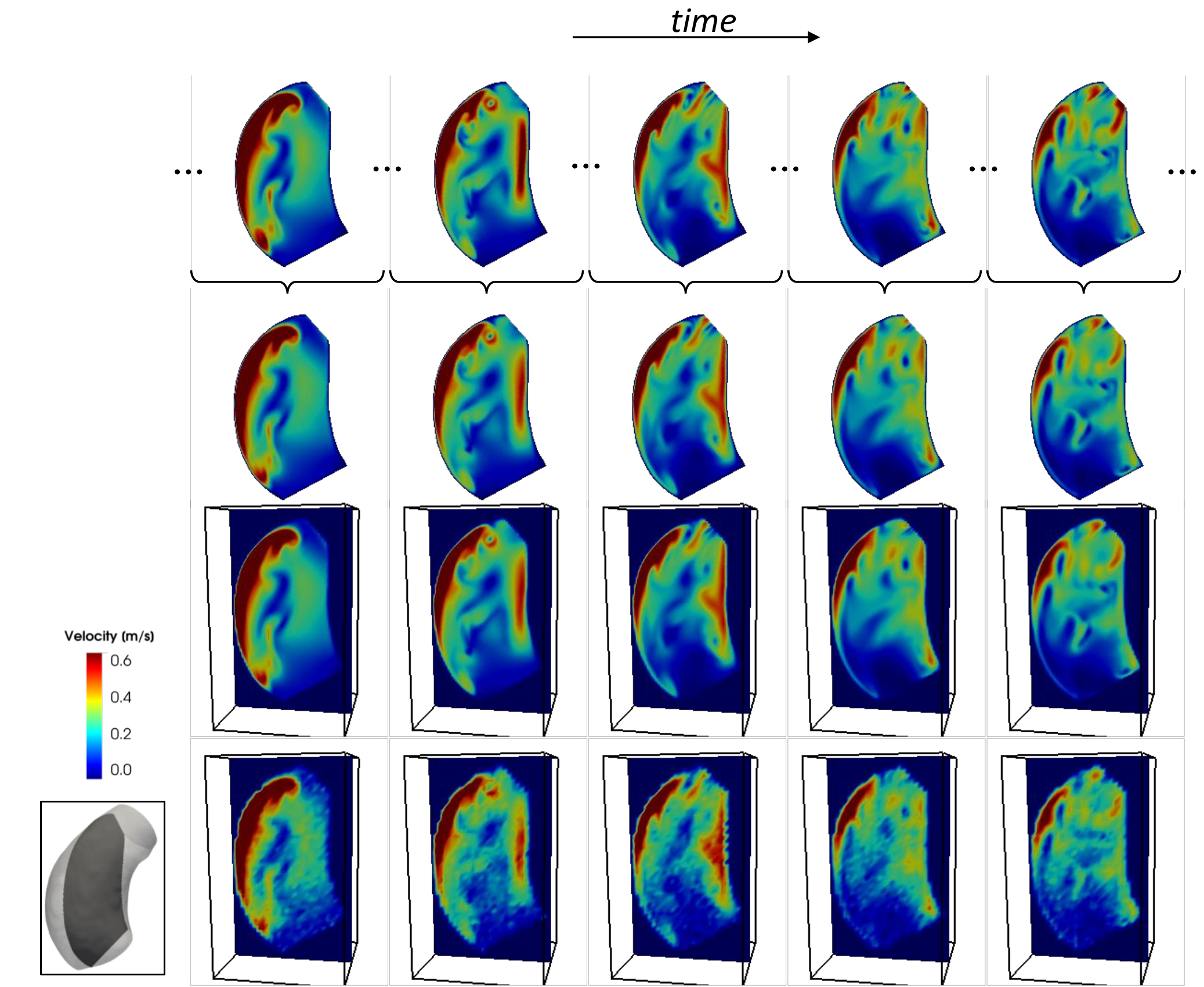}
    \caption{Visualization of velocity magnitude colormaps on a sagittally oriented 2D slice (bottom left) corresponding to the different steps to create synthetic 4D flow data from CFD results. CFD results (top row) are temporally averaged (second row). Temporally average velocity fields defined on unstructured meshes are resampled to a fine 3D Cartesian grid (third row). Fine cartesian grids are downsampled in \textit{k}-space and complex Gaussian noise is added to produce synthetic flow measurments (bottom row).}
    \label{fig:synth_panel_method}
\end{figure}

    \paragraph{Synthetic data creation.}
    Noise-free, high resolution CFD velocity fields were processed to obtain low resolution velocity fields corrupted with noise typical of 4D flow MRI measurements. To achieve this, the following steps were implemented partially following \citep{ferdian20204dflownet}:
    \begin{enumerate}
        \item the sequence of CFD solution snapshots denoted by $\bm{u}^\bullet$ was temporally downsampled to a sequence of $M = \frac{M'}{h}$ velocity fields $\bar{\bm{u}}$, using a moving average such that: $\bar{\bm{u}}(\bm{x}'_n,t'_j) = \frac{1}{h} \sum_{k=1}^{m+h-1} \bm{u}^\bullet(\bm{x}'_n,t'_k)$;
        
        \item each time frame of $\bar{u}$ was converted to a uniform Cartesian grid with voxel size of 1 mm\textsuperscript{3} using a linear interpolation scheme to assign velocity vector values to grid cells, yielding a sequence of Cartesian grids $\tilde{U} [\tilde{\bm{x}}_i, t_j] \in \mathbb{R}^{N_r \times N_c \times N_s \times N_t \times 3}$, with $N_r=76$, $N_c=112$ and $N_s=292$;

        \item\label{step_wrap} each velocity grid in $\tilde{U}$ was converted to a complex tensor containing magnitude and phase images using suitable VENC values as formalized in equation \ref{eq:complex_img};
         
         \item the fast Fourier transform was applied to obtain the corresponding \textit{k}-space data;
         
         \item a truncation of the 3D \textit{k}-space data (high frequencies) was performed, effectively decreasing the spatial resolution by a factor of 2;

         \item a zero-mean Gaussian noise with standard deviation $\sigma$ corresponding to the desired SNR (calculated according to \citep{ferdian20204dflownet}) was added to the \textit{k}-space data;

         \item a randomized sampling mask drawn from a normal distribution and covering S\% percent of the \textit{k}-space was applied to further undersample the frequency content, keeping a fully sampled calibration region of 5 $\times$ 5 $\times$ 5 in the center of \textit{k}-space;

         \item the inverse Fourier transform was applied to the undersampled, noise-corrupted \textit{k}-space, yielding a complex tensor of magnitude and phase images;
         
         \item complex images were converted back to real images of velocity fields using VENC values consisted with step \ref{step_wrap}, obtaining a sequence of noisy synthetic velocity measurements $U^*$ defined at voxel coordinates $\bm{x}_i, i=1, ..., N$ with isotropic voxel size of 2$\times$2$\times$2 mm\textsuperscript{3} and at times $t_j, j=1,...,M$.

    \end{enumerate}
For each velocity direction, the VENC value was chosen 10\% larger than the maximum velocity, so that the phase wrapping/unwrapping would not introduce aliasing artifacts.
Figure \ref{fig:synth_panel_method} shows the effect of the different implemented steps to create our synthetic 4D flow data.

The level of degradation of our synthetic measurements is regulated by three parameters: $h$, SNR and $S$. We chose $h=40$ to obtain $\Delta t = 0.04 s$, which is compatible with real medical measurements. 
We tested our approach on three different levels of degradation, \textit{mild}, \textit{medium} or \textit{extreme} reported in Table \ref{tab:noise_level} and shown in Figure \ref{fig:noise_levels_method}.

\begin{table}[htbp]
\centering
\begin{tabular}{lcc}
\hline
Noise level & SNR & $S$\\ \hline
\textit{mild}    &  20 &  99  \\
\textit{medium}  &  5  &  95  \\
\textit{extreme}     &  2  &  68  \\
\hline
\end{tabular}
\caption{Degradation parameters for generating synthetic measurements from the reference velocity field.}
\label{tab:noise_level}
\end{table}

\subsubsection{Comparison with existing methods} \label{sec:existing_methods}
The denoising and SR performances of our approach were compared against existing methods. Among them, only 4D RBFs lend themselves to the joint task of denoising and SR. Hence, pure denoising approaches were combined with simple interpolation schemes to achieve SR. The following methods were tested:
\begin{enumerate}
    \item linear interpolation (LITP) as a baseline for SR;
    
    \item DFW with automatic threshold selection based on \textit{SureShrink} \citep{ong2015robust} for denoising and LITP for SR. Velocity fields generated by this approach will be denoted as DFW;

    \item 3D DF-RBFs as described in \citep{busch2013reconstruction} for denoising and LITP for SR. Velocity fields generated by this approach will be denoted as DF-RBF;

    \item an approach based on 4D RBFs \citep{thewlis20234}, but for which no official implementation was available. Hence, we implemented our version of 4D RBFs with multi-quadric kernel and local support for denoising and SR. We set 10 as the number of nearest neighboring points for local kernel support. A \textit{soft} enforcement of the no-slip condition on the vessel wall was applied by setting null velocity values at point belonging to the wall region. Velocity fields generated by this approach will be denoted as 4D-RBF.      
\end{enumerate}

\subsection{Case 2: \textit{in vivo} 4D flow MRI}
\begin{figure}[htbp]
    \centering
    \includegraphics[width=0.6\textwidth]{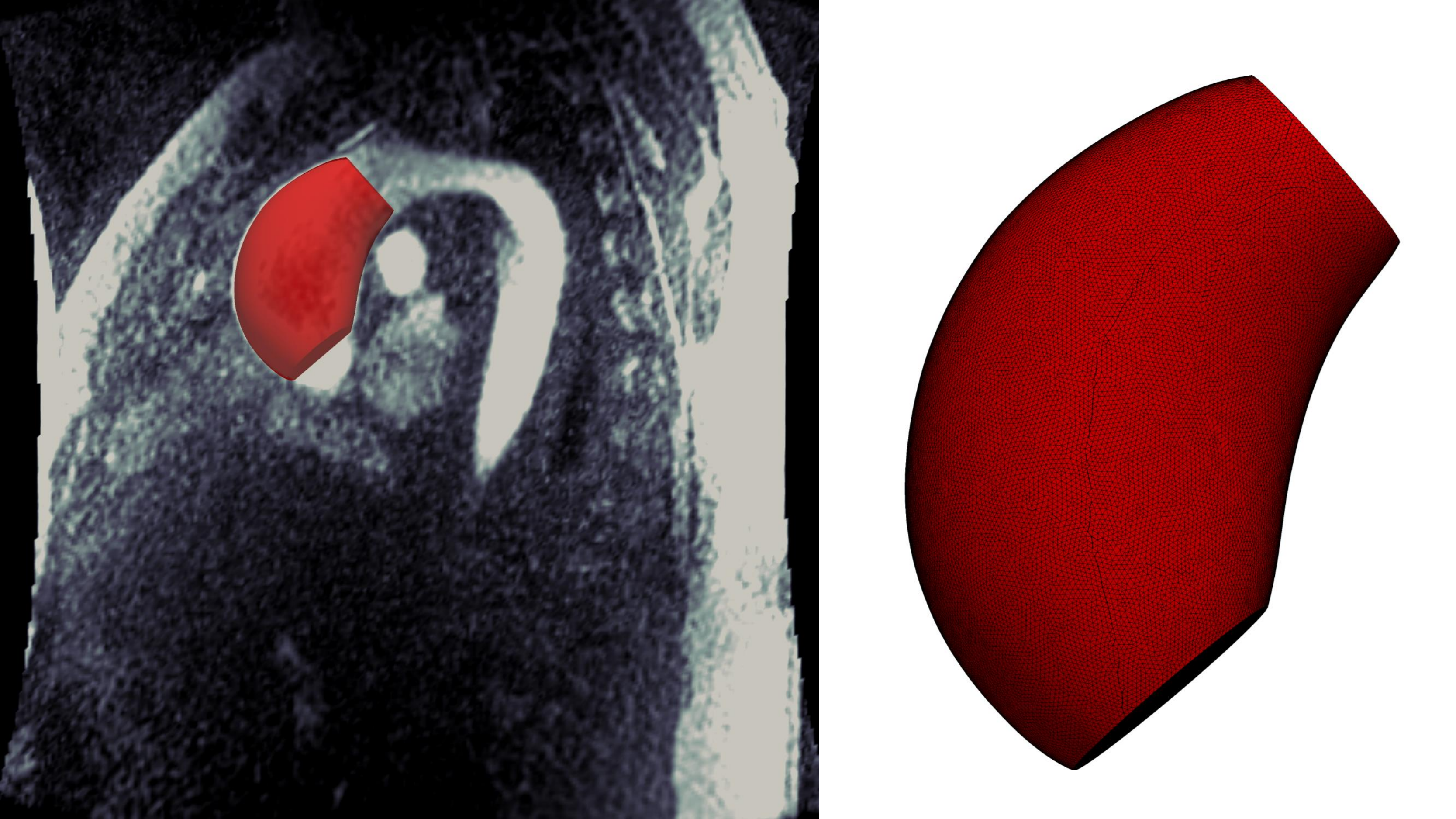}
    \caption{Left: segmented 3D geometry (red) superimposed on a slice representation of PCMRA images. Right: fine 3D mesh for spatial super-resolution}
    \label{fig:pcmra_seg}
\end{figure}

A thoracic 4D flow MRI scan of a subject with ascending thoracic aortic aneurysm was retrospectively retrieved. Images were fully deintentified and provided by Weill Cornell Medicine, (NY, USA). A respiratory compensated technique was adopted with the following settings: spatial resolution (voxel size) 1.14 mm $\times$ 1.14 mm $\times$ 0.9 mm, field of view = 360 mm, flip angle = 15°, VENC = 200 cm/s in all 3 directions, time between consecutive frames = 30 ms, for a total of 20 frames per cardiac cycle.
DICOM images were processed using open-source code \citep{saitta2022data} to compute the PCMRA image and extract the segmentation of the enlarged ascending aortic tract (Figure \ref{fig:pcmra_seg}).

\begin{figure}[htbp]
    \centering
    \includegraphics[width=0.99\textwidth]{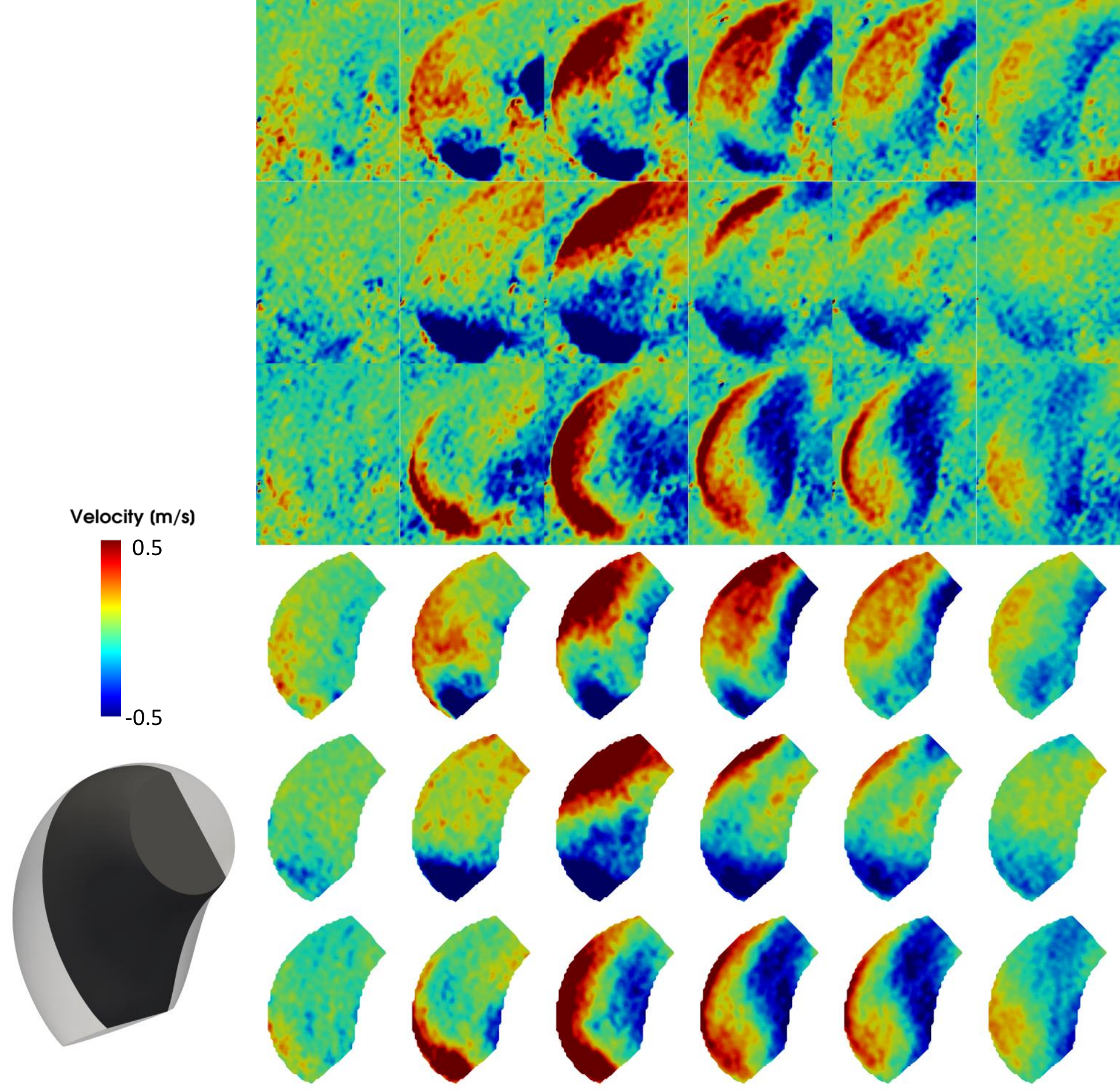}
    \caption{Sagittally oriented 2D slice within the aortic aneurysm (bottom left), together with colormaps of left-to-right (first and fourth rows), posterior-to-anterior (second and fifth rows) and foot-to-head (third and sixth rows) velocity components. Rows 4 to 6 show velocities sampled on a 3D mesh by linear interpolation. Columns from left to right correspond to increasing time points.}
    \label{fig:phases_sagittal}
\end{figure}

\begin{figure}[htbp]
    \centering
    \includegraphics[width=0.99\textwidth]{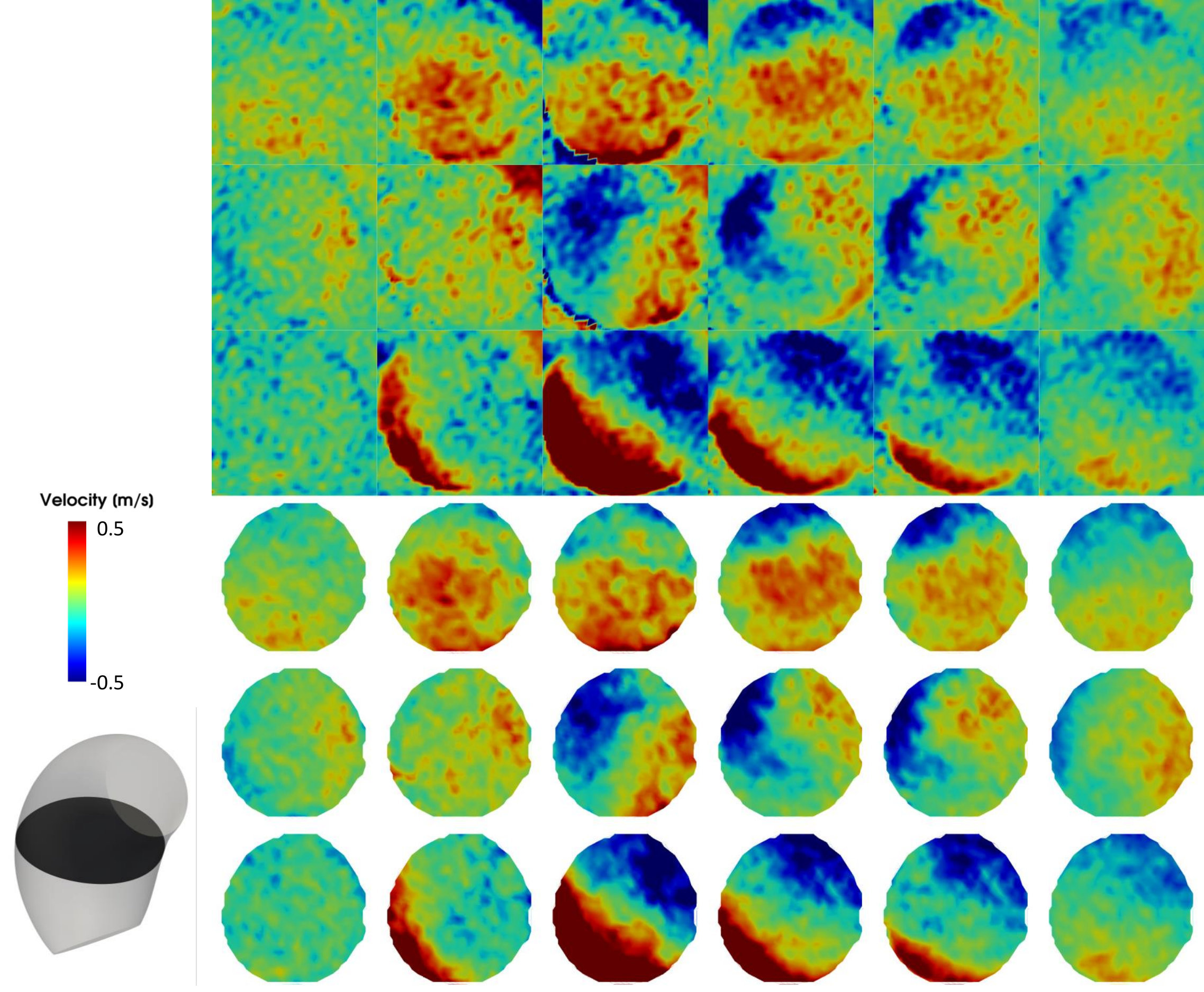}
    \caption{Axially oriented 2D slice within the aortic aneurysm (bottom left), together with colormaps of left-to-right (first and fourth rows), posterior-to-anterior (second and fifth rows) and foot-to-head (third and sixth rows) velocity components. Rows 4 to 6 show velocities sampled on a 3D mesh by linear interpolation. Columns from left to right correspond to increasing time points.}
    \label{fig:phases_axial}
\end{figure}

The velocity components measured through flow MRI can be visualized in Figures \ref{fig:phases_sagittal} and \ref{fig:phases_axial} for a sagittally and an axially oriented 2D slice, respectively.

\FloatBarrier

\section{Results}
\subsection{Case 1: quantification of image degradation}
\begin{figure}[htbp]
    \centering
    \includegraphics[width=\textwidth]{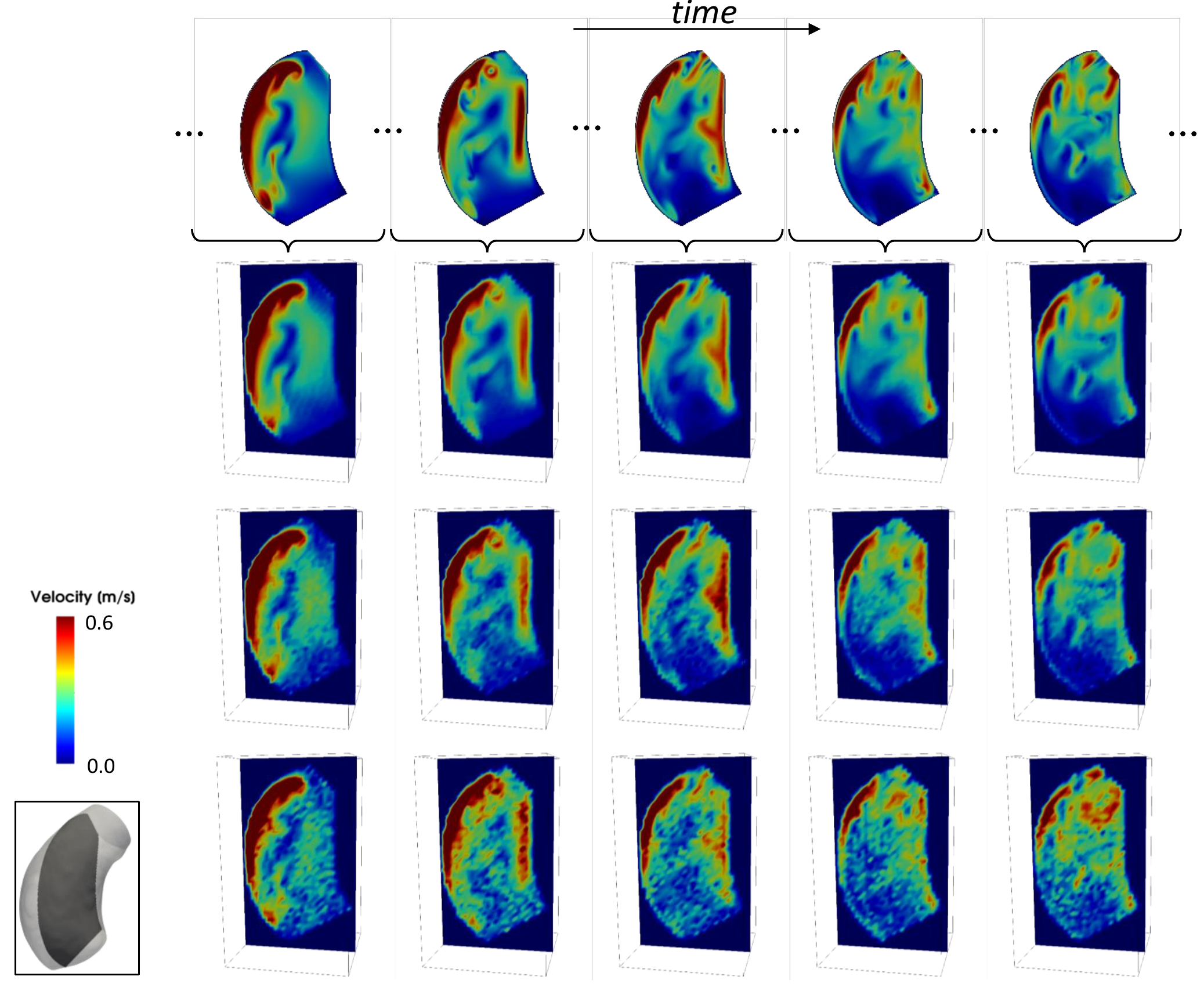}
    \caption{Velocity magnitude colormaps on a sagittal slice (bottom left corner) corresponding to: CFD solution (top row), images with \textit{mild} added noise (second row), images with \textit{medium} added noise (third row), images with \textit{extreme} added noise (bottom row). Columns from left to right correspond to increasing time points.}
    \label{fig:noise_levels_method}
\end{figure}

\begin{figure}[htbp]
    \centering
    \includegraphics[width=\textwidth]{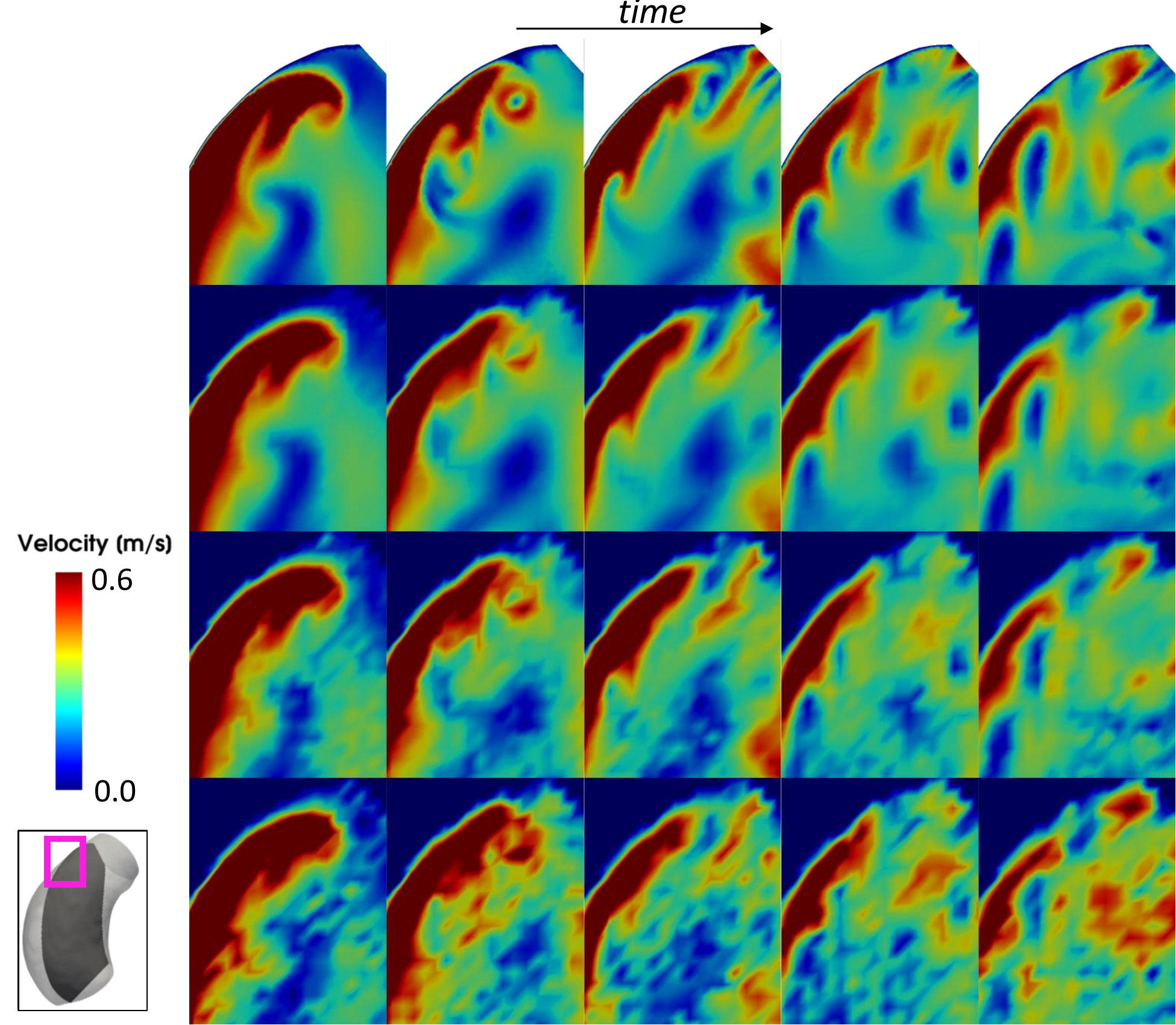}
    \caption{Detailed view of velocity magnitude colormaps on a sagittal slice (magenta rectangle in bottom left corner) corresponding to: CFD solution (top row), images with \textit{mild} added noise (second row), images with \textit{medium} added noise (third row), images with \textit{extreme} added noise (bottom row).}
    \label{fig:noise_levels_detail}
\end{figure}

The implemented degradation process to transform CFD velocity fields into synthetic flow MR measurements consisted of a sequence of steps that gradually worsened the quality of the data. The contribution of each step described in Section \ref{sec:synth_method} is reported in Table \ref{tab:noise_errors}. To make these comparisons, the CFD velocity fields were evaluated at coarse spatio-temporal discretizations as denoted in Section \ref{sec:synth_method}.
Qualitatively, the addition of \textit{mild} complex noise produced velocity fields that well kept the low frequency features present in the original fields (Figures \ref{fig:noise_levels_method} and \ref{fig:noise_levels_detail}, rows 1 and 2). Synthetic images with \textit{medium} noise level produced more degraded images, with clearly visible artifacts and loss of high frequency details (Figures \ref{fig:noise_levels_method} and \ref{fig:noise_levels_detail}, rows 1 and 3). The \textit{extreme} noise level yielded severely worsened velocity field with respect to ground truth, with a visible complete loss of fine flow details and generally higher velocities (Figures \ref{fig:noise_levels_method} and \ref{fig:noise_levels_detail}, rows 1 and 4).

\begin{table}[htbp]
\centering
\begin{tabular}{lccc}
\hline
\multicolumn{1}{c}{} & \multicolumn{1}{l}{$\bm{u}_{ref} = \bm{u}^\bullet (\bm{x}', t)$} & $\bm{u}_{ref} = \bm{u}^\bullet (\tilde{\bm{x}}, t)$ & $\bm{u}_{ref} = \bm{u}^\bullet (\bm{x}, t)$ \\ 
                       & $\bm{u} = \bar{\bm{u}} (\bm{x}', t)$ & $\bm{u} = \tilde{\bm{U}} [\tilde{\bm{x}}, t]$ & \multicolumn{1}{c}{$\bm{u} = \bm{U}^* [\bm{x}, t]$} \\ \hline
\multicolumn{1}{c}{\multirow{3}{*}{mNRMSE {[}\%{]}}} & \multirow{3}{*}{1.79}         & \multirow{3}{*}{1.75} &  3.63    \\ 
\multicolumn{1}{c}{} &   &   &            5.24            \\ 
\multicolumn{1}{c}{} &   &   &            8.23            \\[0.2cm]
\multirow{3}{*}{vNMRSE{[}\%{]}}                        & \multirow{3}{*}{2.06}         & \multirow{3}{*}{2.01} & 3.22   \\  
                       &   &   &       5.60                 \\  
                       &   &   &         9.64               \\ [0.2cm]
\multirow{3}{*}{DE {[}\%{]}}                             & \multirow{3}{*}{4.95}         & \multirow{3}{*}{0.59} &  1.41    \\  
                       &   &   &         7.96               \\ 
                       &   &   &          15.9              \\ \hline
\end{tabular}
\caption{Errors introduced by the implemented degradation steps.}
\label{tab:noise_errors}
\end{table}

\subsection{Case 1: hyperparameter tuning}
\begin{figure}[htbp]
    \centering
    \includegraphics[width=\textwidth]{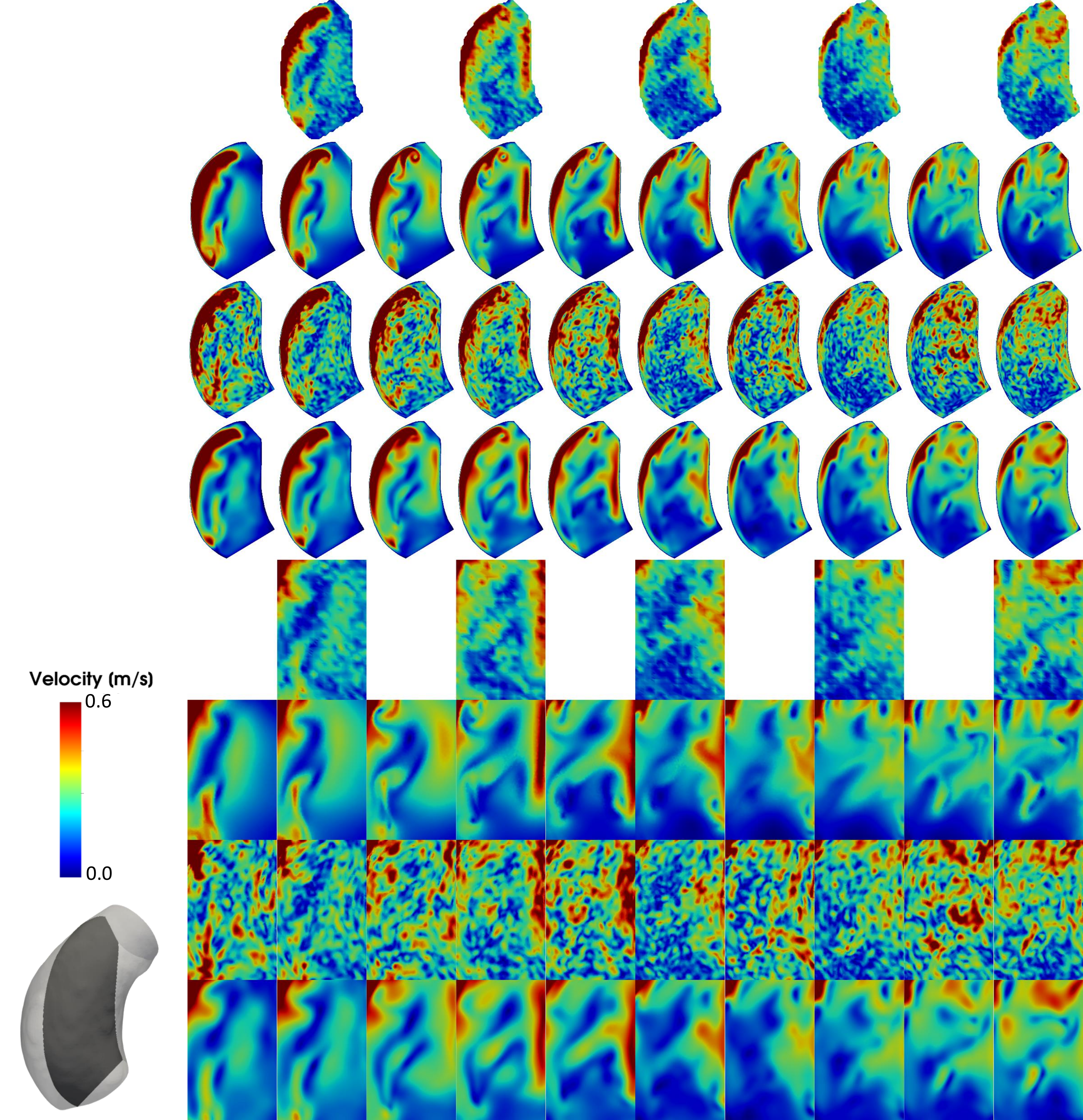}
    \caption{Sagittal view of the effect of different SIREN architectures on synthetic images with \textit{extreme} noise level. }
    \label{fig:overfit_extreme_sagittal}
\end{figure}

\begin{figure}[htbp]
    \centering
    \includegraphics[width=\textwidth]{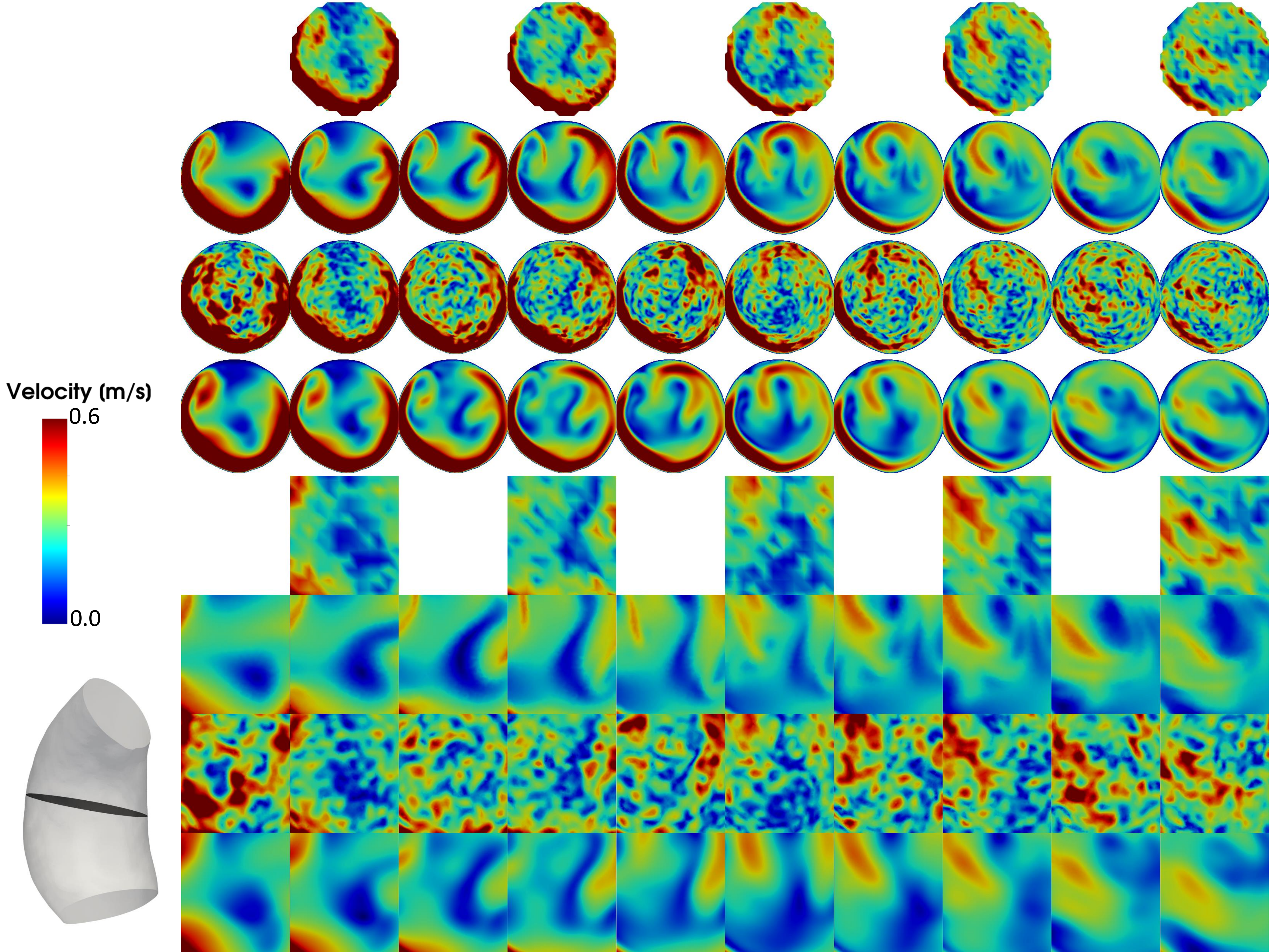}
    \caption{Axial view of the effect of different SIREN architectures on synthetic images with \textit{extreme} noise level. }
    \label{fig:overfit_extreme_axial}
\end{figure}

The effects of \textsc{siren}'s depth (number of layers) and width (number of neurons per layer) on the denoising and SR performances were assessed by training different configurations of $f_\mathbf{\Theta}$ on synthetic images with \textit{mild}, \textit{medium} and \textit{extreme} noise level. Trained models were evaluated on CFD nodal mesh coordinates $\bm{x}'$ and at time points $t'$ evenly spaced by 0.004 s, effectively oversampling $\Omega$ by $\approx \times 20$ and the time interval [0.2 - 0.399] by $\times10$. Results are reported in Tables \ref{tab:hyperparams_results_mild}, \ref{tab:hyperparams_results_medium} and \ref{tab:hyperparams_results_extreme}, for \textit{mild}, \textit{medium} and \textit{extreme} noise levels, respectively.
For \textit{mild} noise levels, all the tested \textsc{siren} configurations gave low errors with respect to ground truth velocity fields. Networks with greater width and depth resulted in only slightly lower mNRMSE, vNRMSE and DE. 
For \textit{medium} noise levels, better results were obtained by wider models, but not necessarily by deeper ones. For this noise settings, the best performing \textsc{siren} consisted of 12 layers, each with 500 neurons.
In the case of \textit{extreme} noise, wider networks gave worse results than narrower ones, while deeper architectures generally produced more accurate velocity fields compared to ground truth. Wider models showed a tendency to overfit high frequency noise, as shown in Figures \ref{fig:overfit_extreme_sagittal} and \ref{fig:overfit_extreme_axial}. \\
We chose the single best configuration as the one that minimized the sum of vNRMSE, mNRMSE and DE for all three noise levels, hence a \textsc{siren} 20 layer deep and with 300 neurons per layer was selected for all downstream comparisons against other methods and on the medical dataset.

\begin{table}[htbp]
\centering
\begin{tabular}{llllllllllll}
\hline
\multicolumn{7}{c}{\emph{Depth}}                               \\ 
\parbox[t]{2mm}{\multirow{17}{*}{\rotatebox[origin=c]{90}{\emph{Width}}}} &
  \multicolumn{1}{c}{} &
  4 &
  \multicolumn{1}{c}{8} &
  \multicolumn{1}{c}{12} &
  \multicolumn{1}{c}{16} &
  \multicolumn{1}{c}{20} \\ \hline 
 & \multicolumn{1}{c}{}     & 3.52 & 3.56 & 3.53 & 3.42 & 3.46  \\
 & \multicolumn{1}{c}{100}  & 3.43 & 3.36 & 3.32 & 3.22 & 3.27 \\
 & \multicolumn{1}{c}{}     & 6.8 & 6.4 & 6.3 & 6.22 & 6.31 \\[0.2cm]
 
 &                         & 3.48 & 3.46 & 3.52 & 3.52 & 3.41\\
 & \multicolumn{1}{c}{200} & 3.36 & 3.27 & 3.27 & 3.27 & 3.21 \\
 &                         & 6.43 & 6.23 & 6.16 & 6.16 & 6.14 \\[0.2cm]
 
 &                         & 3.51 & 3.53 & 3.62 & 3.50 & 3.43  \\
 & \multicolumn{1}{c}{300} & 3.40 & 3.30 & 3.33 & 3.24 & 3.21   \\
 &                         & 6.47 & 6.24 & 6.19 & 6.14 & 6.13  \\[0.2cm]
 
 &                         & 3.49 & 3.47 & 3.60 & 3.49 & 3.53 \\
 & \multicolumn{1}{c}{400} & 3.34 & 3.25 & 3.31 & 3.25 & 3.29  \\
 &                         & 6.37 & 6.16 & 6.17 & 6.12 & 6.15 \\[0.2cm]
 
 &                         & 3.5.9 & 3.47 & 3.55 & 35.4 & 34.5  \\
 & \multicolumn{1}{c}{500} & 3.4.5 & 3.26 & 3.30 & 32.9 & 32.3 \\
 &                         & 6.42 & 6.19 & 6.17 & 6.19 & 6.14 \\ \hline

\end{tabular}
\caption{Effect of \textsc{siren} number of layers (depth) and number of neurons per layer (width) on denoising and super-resolution using data with \textit{mild} noise level. In each cell values of mNRMSE (top), vNRMSE (middle) and DE (bottom) are reported.}
\label{tab:hyperparams_results_mild}
\end{table}

\begin{table}[htbp]
\centering
\begin{tabular}{llllllllllll}
\hline
\multicolumn{7}{c}{\emph{Depth}}                               \\ 
\parbox[t]{2mm}{\multirow{17}{*}{\rotatebox[origin=c]{90}{\emph{Width}}}} &
  \multicolumn{1}{c}{} &
  4 &
  \multicolumn{1}{c}{8} &
  \multicolumn{1}{c}{12} &
  \multicolumn{1}{c}{16} &
  \multicolumn{1}{c}{20} \\ \hline 
 & \multicolumn{1}{c}{}     & 4.48 & 4.63 & 4.65 & 5.86 & 5.29  \\
 & \multicolumn{1}{c}{100}  & 4.73 & 4.92 & 4.94 & 6.13 & 5.61 \\
 & \multicolumn{1}{c}{}     & 10.5 & 10.8 & 10.9 & 13.6 & 12.3 \\[0.2cm]
 
 &                         & 3.98 & 3.98 & 4.42 & 4.49 & 4.85\\
 & \multicolumn{1}{c}{200} & 4.24 & 4.24 & 4.72 & 4.81 & 5.15 \\
 &                         & 9.66 & 9.61 & 10.5 & 10.7 & 11.5 \\[0.2cm]
 
 &                         & 3.94 & 3.94 & 3.97 & 4.02 & 4.20  \\
 & \multicolumn{1}{c}{300} & 4.20 & 4.20 & 4.21 & 4.30 & 4.51   \\
 &                         & 9.59 & 9.56 & 9.5 & 9.67 & 10.21  \\[0.2cm]
 
 &                         & 4.03 & 3.81 & 3.89 & 3.87 & 4.03 \\
 & \multicolumn{1}{c}{400} & 4.35 & 4.09 & 4.16 & 4.12 & 4.28  \\
 &                         & 9.87 & 9.39 & 9.48 & 9.44 & 9.69 \\[0.2cm]
 
 &                         & 3.95 & 3.82 & 3.83 & 3.85 & 3.84  \\
 & \multicolumn{1}{c}{500} & 4.24 & 4.11 & 4.08 & 4.10 & 4.10 \\
 &                         & 9.62 & 9.41 & 9.38 & 9.33 & 9.36 \\ \hline

\end{tabular}
\caption{Effect of \textsc{siren} number of layers (depth) and number of neurons per layer (width) on denoising and super-resolution using data with \textit{medium} noise level. In each cell values of mNRMSE (top), vNRMSE (middle) and DE (bottom) are reported.}
\label{tab:hyperparams_results_medium}
\end{table}

\begin{table}[htbp]
\centering
\begin{tabular}{llllllllllll}
\hline
\multicolumn{7}{c}{\emph{Depth}}                               \\ 
\parbox[t]{2mm}{\multirow{17}{*}{\rotatebox[origin=c]{90}{\emph{Width}}}} &
  \multicolumn{1}{c}{} &
  4 &
  \multicolumn{1}{c}{8} &
  \multicolumn{1}{c}{12} &
  \multicolumn{1}{c}{16} &
  \multicolumn{1}{c}{20} \\ \hline 
 & \multicolumn{1}{c}{}     & 5.31 & 5.37 & 5.26 & 5.30 & 5.87  \\
 & \multicolumn{1}{c}{100}  & 5.73 & 5.80 & 5.73 & 5.75 & 6.29 \\
 & \multicolumn{1}{c}{}     & 13.01 & 13.33 & 12.93 & 13.02 & 14.17 \\[0.2cm]
 
 &                         & 5.84 & 5.99 & 5.16 & 5.15 & 5.21\\
 & \multicolumn{1}{c}{200} & 6.62 & 6.91 & 5.68 & 5.67 & 5.64 \\
 &                         & 14.83 & 15.11 & 12.87 & 12.78 & 12.84 \\[0.2cm]
 
 &                         & 7.30 & 7.74 & 5.40 & 5.21 & 5.16  \\
 & \multicolumn{1}{c}{300} & 8.50 & 9.05 & 6.09 & 5.77 & 5.63   \\
 &                         & 18.86 & 19.91 & 13.41 & 12.95 & 12.7  \\[0.2cm]
 
 &                         & 8.86 & 8.14 & 7.61 & 5.56 & 5.44 \\
 & \multicolumn{1}{c}{400} & 10.3 & 9.52 & 8.89 & 6.33 & 6.10  \\
 &                         & 22.34 & 21.02 & 19.68 & 13.98 & 13.51 \\[0.2cm]
 
 &                         & 9.41 & 8.07 & 7.95 & 7.46 & 6.50  \\
 & \multicolumn{1}{c}{500} & 10.9 & 9.45 & 9.28 & 8.65 & 7.56 \\
 &                         & 23.52 & 20.88 & 20.5 & 19.16 & 16.66 \\ \hline

\end{tabular}
\caption{Effect of \textsc{siren} number of layers (depth) and number of neurons per layer (width) on denoising and super-resolution using data with \textit{extreme} noise level. In each cell values of mNRMSE (top), vNRMSE (middle) and DE (bottom) are reported.}
\label{tab:hyperparams_results_extreme}
\end{table}

\FloatBarrier

\subsection{Case 1: comparison with existing methods}
\subsubsection{Velocity fields}
\begin{figure}[htbp]
    \centering
    \includegraphics[width=\textwidth]{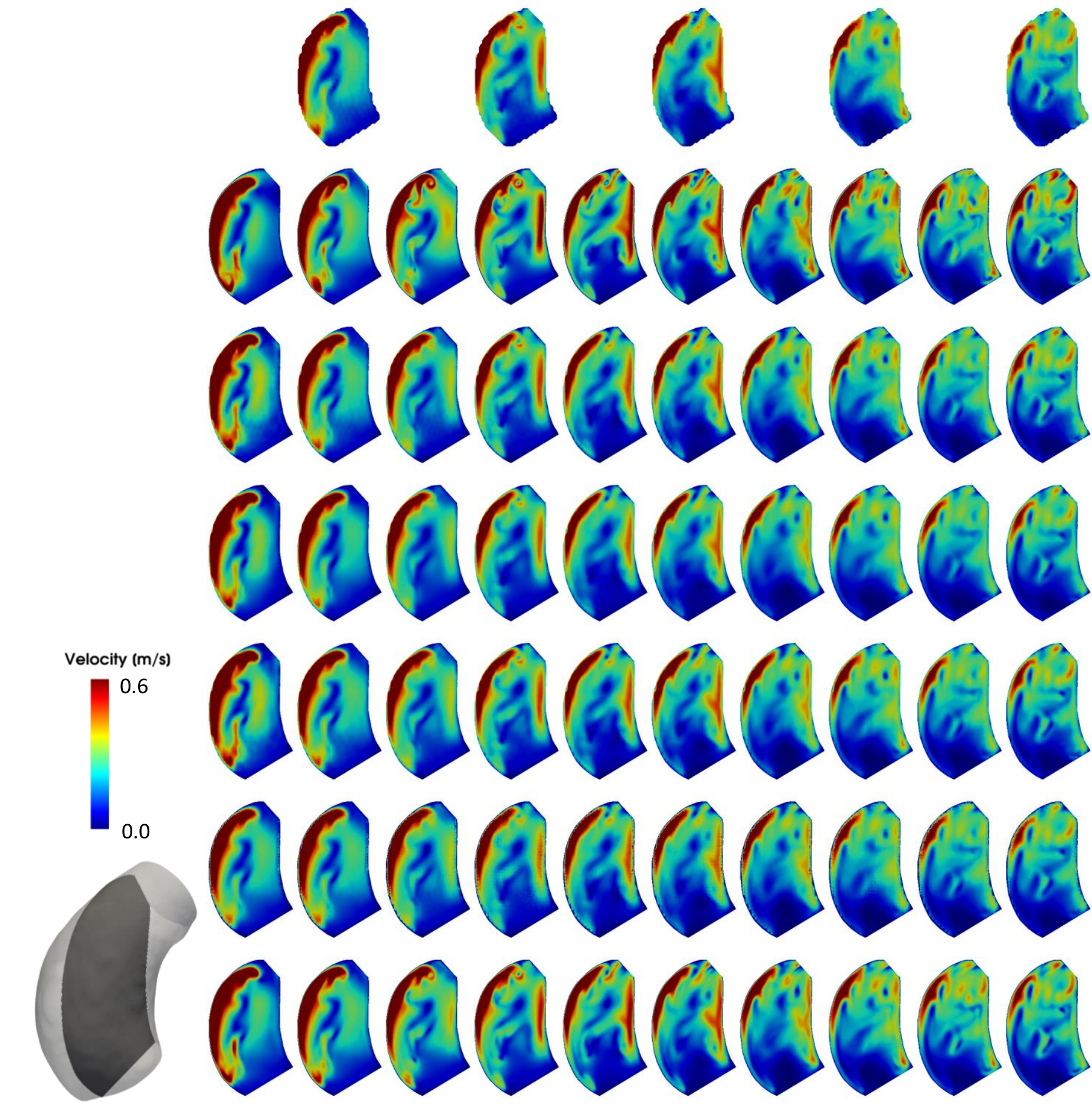}
    \caption{Velocity colormaps on a 2D sagittal slice (bottom left). Top row: synthetic measurements with \textit{mild} noise level at time points evenly spaced by $\Delta t = 40 ms$. Second row: ground truth velocity fields from CFD at time points evenly spaced by $\Delta t = 20 ms$. Third row: LITP results. Fourth row: DFW results. Fifth row: DF-RBF results. Sixth row: 4D-RBF results. Seventh row: our method.}
    \label{fig:compare_vel_mild_sag}
\end{figure}

\begin{figure}[htbp]
    \centering
    \includegraphics[width=\textwidth]{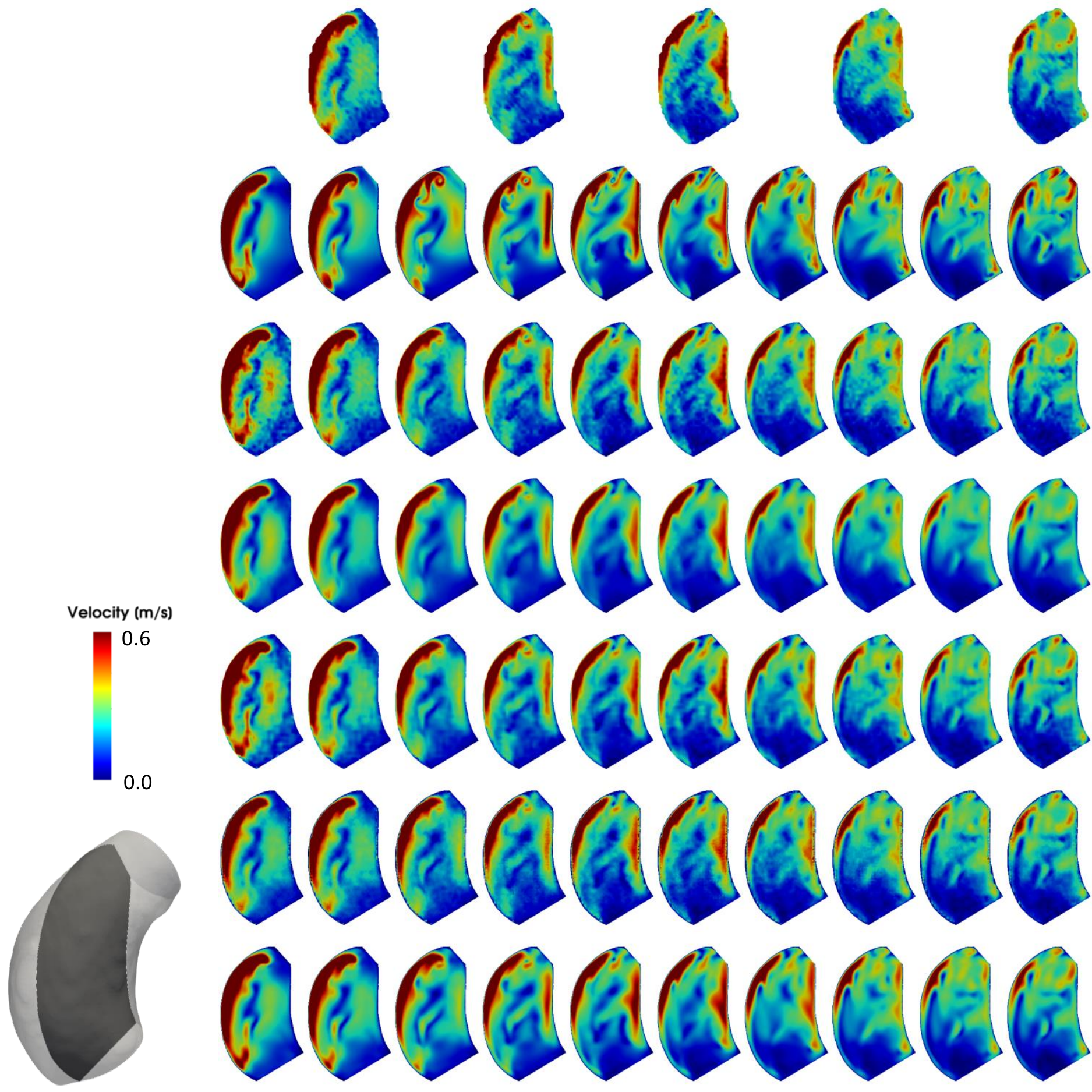}
    \caption{Velocity colormaps on a 2D sagittal slice (bottom left). Top row: synthetic measurements with \textit{medium} noise level at time points evenly spaced by $\Delta t = 40 ms$. Second row: ground truth velocity fields from CFD at time points evenly spaced by $\Delta t = 20 ms$. Third row: LITP results. Fourth row: DFW results. Fifth row: DF-RBF results. Sixth row: 4D-RBF results. Seventh row: our method.}
    \label{fig:compare_vel_medium_sag}
\end{figure}

\begin{figure}[htbp]
    \centering
    \includegraphics[width=\textwidth]{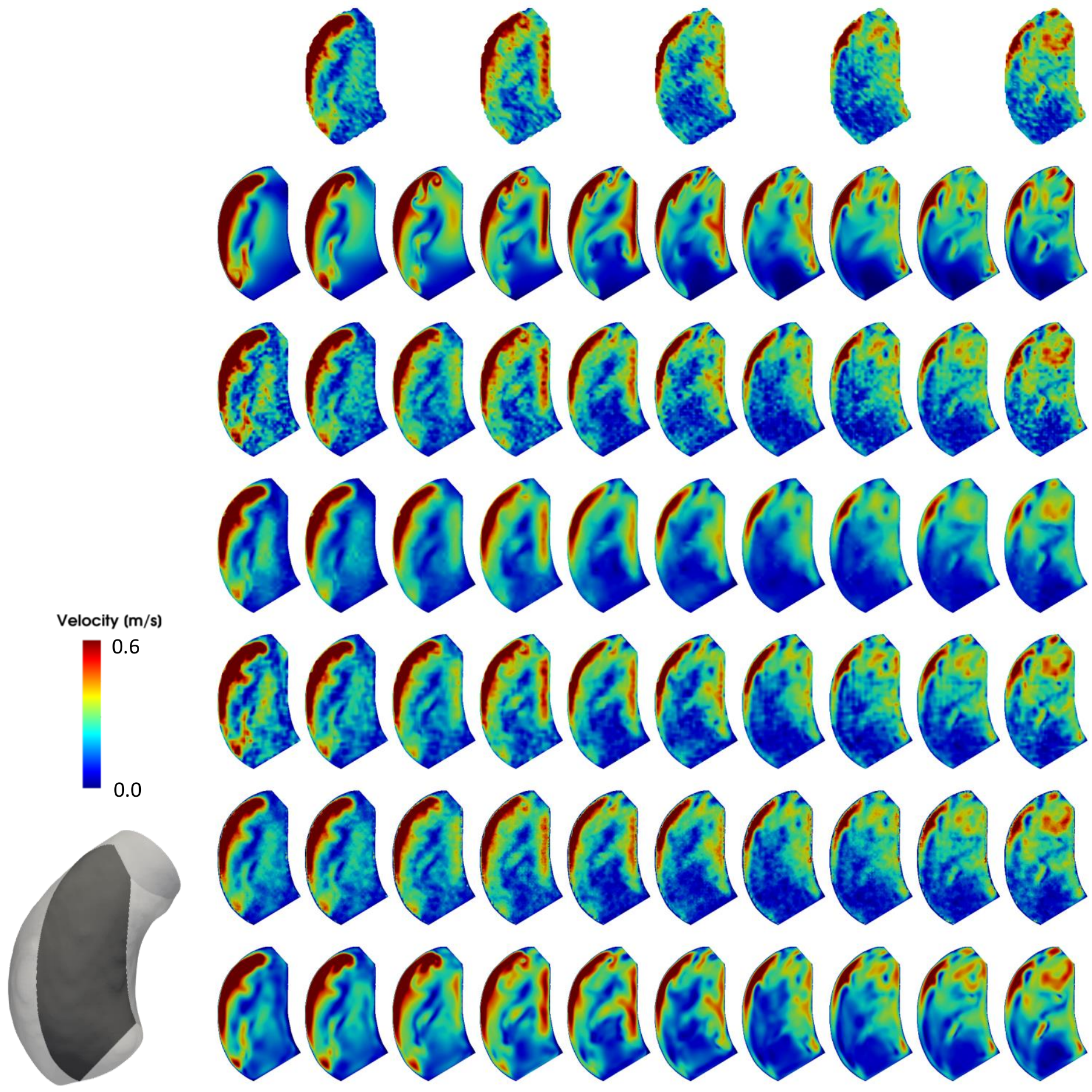}
    \caption{Velocity colormaps on a 2D sagittal slice (bottom left). Top row: synthetic measurements with \textit{extreme} noise level at time points evenly spaced by $\Delta t = 40 ms$. Second row: ground truth velocity fields from CFD at time points evenly spaced by $\Delta t = 20 ms$. Third row: LITP results. Fourth row: DFW results. Fifth row: DF-RBF results. Sixth row: 4D-RBF results. Seventh row: our method.}
    \label{fig:compare_vel_extreme_sag}
\end{figure}

The \textsc{siren} configuration that showed the best results was evaluated on the synthetic images with three different levels of noise. Velocity fields obtained with our approach were compared against the existing methods listed in Section \ref{sec:existing_methods}. The proposed \textsc{siren} gave lower mNRMSE, vNRMSE and DE for all noise levels (Table \ref{tab:comparison_results_vel}). Among the LITP-based methods, DF-RBF \citep{busch2013reconstruction} performed better than LITP alone, while DFW provided the worst performance. 
Qualitatively, all existing methods produced very similar velocity fields for the \textit{mild} noise level case, showing suboptimal capabilites of reconstructing finer flow details (Figure \ref{fig:compare_vel_mild_sag}). For the same level of noise, our method was able to fit flow details more accurately. 
Similar results were observed for the \textit{medium} and \textit{extreme} noise levels. As shown in Figures \ref{fig:compare_vel_medium_sag} and \ref{fig:compare_vel_extreme_sag}, all baseline approaches tended to perform denosing by oversmoothing the data. On the other hand, our method was able to filter out noise but better preserving finer flow structures.

\begin{table}[htbp]
\centering
\begin{tabular}{cccccc}
\hline
 & LITP & \begin{tabular}[c]{@{}c@{}}DFW\\ +\\ LITP\end{tabular} & \begin{tabular}[c]{@{}c@{}}DF-RBF \\ +\\ LITP\end{tabular} & 4DRBF & \textsc{siren} \\ \hline
        & 5.88  & 6.47  & 5.90  & 6.91  & 3.50  \\
\textit{mild}    & 4.96  & 5.40  & 4.99  & 5.68  & 3.24  \\
        & 7.13  & 7.96  & 7.2   & 8.15  & 6.14  \\[0.2cm]
        & 6.11  & 7.26  & 6.05  & 6.79  & 4.02  \\
\textit{medium}  & 5.50  & 6.18  & 5.42  & 6.01  & 4.30  \\
        & 10.1  & 10.16 & 9.78  & 10.53 & 9.67  \\[0.2cm]
        & 7.06  & 8.53  & 6.85  & 7.57  & 5.21  \\
\textit{extreme} & 6.96  & 7.43  & 6.62  & 7.04  & 5.77  \\
        & 14.16 & 13.01 & 13.26 & 13.9  & 12.95 \\ \hline
\end{tabular}
\caption{Velocity field comparison with existing methods. In each cell values of mNRMSE (top), mNRMSE (middle) and DE (bottom) are reported.}
\label{tab:comparison_results_vel}
\end{table}

\subsubsection{Wall shear stress fields}
\begin{figure}[htbp]
    \centering
    \includegraphics[width=\textwidth]{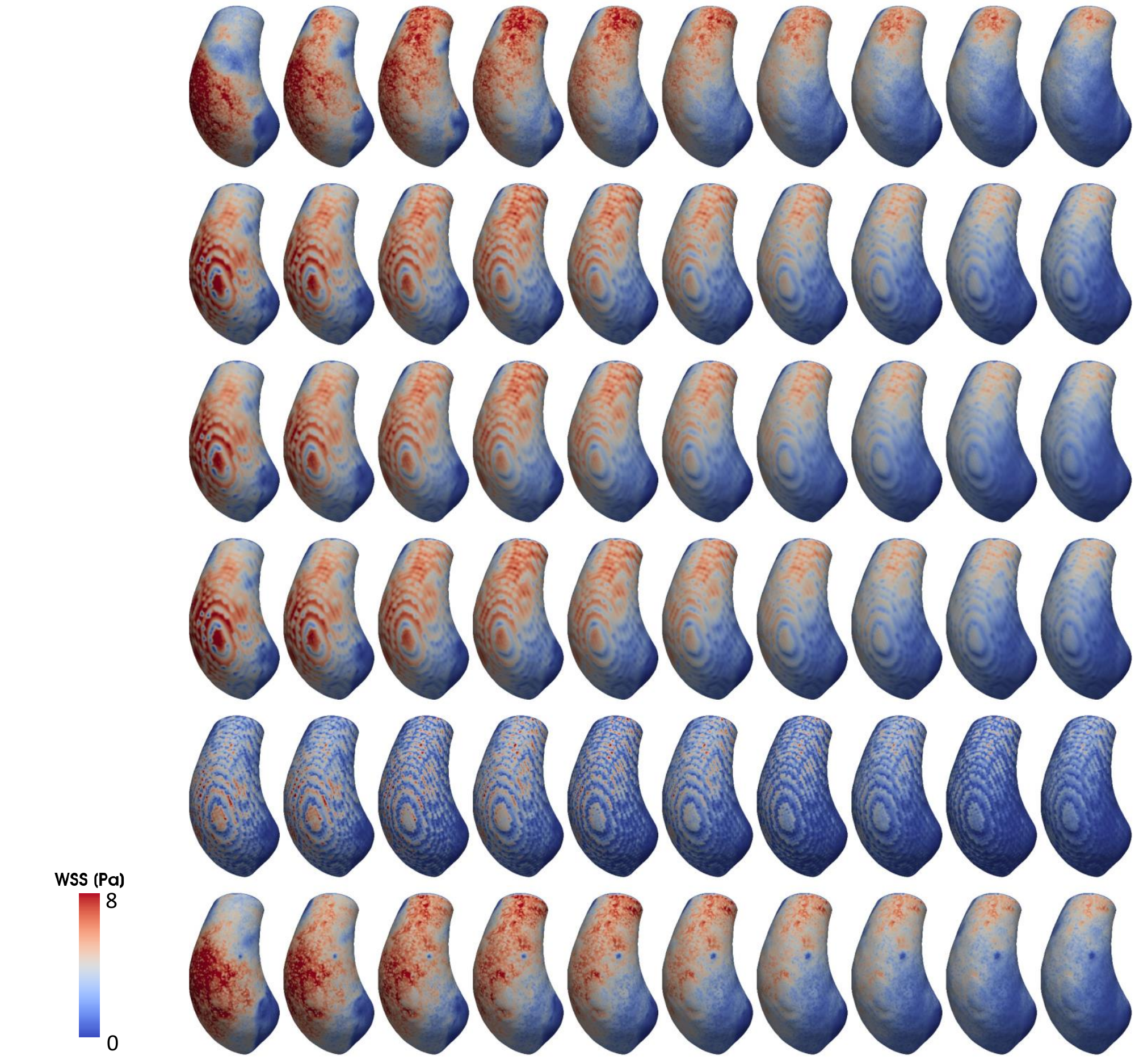}
    \caption{Results corresponding to \textit{mild} noise levels. Wall shear stress (WSS) colormaps on the aortic wall at time points evenly spaced by $\Delta t = 20 ms$. Top row: WSS computed on ground truth velocity fields from CFD. LITP (second row), DFW (third row), DF-RBF (fourth row), 4D-RBF (fifth row), our method (sixth row). }
    \label{fig:compare_wss_mild}
\end{figure}

\begin{figure}[htbp]
    \centering
    \includegraphics[width=\textwidth]{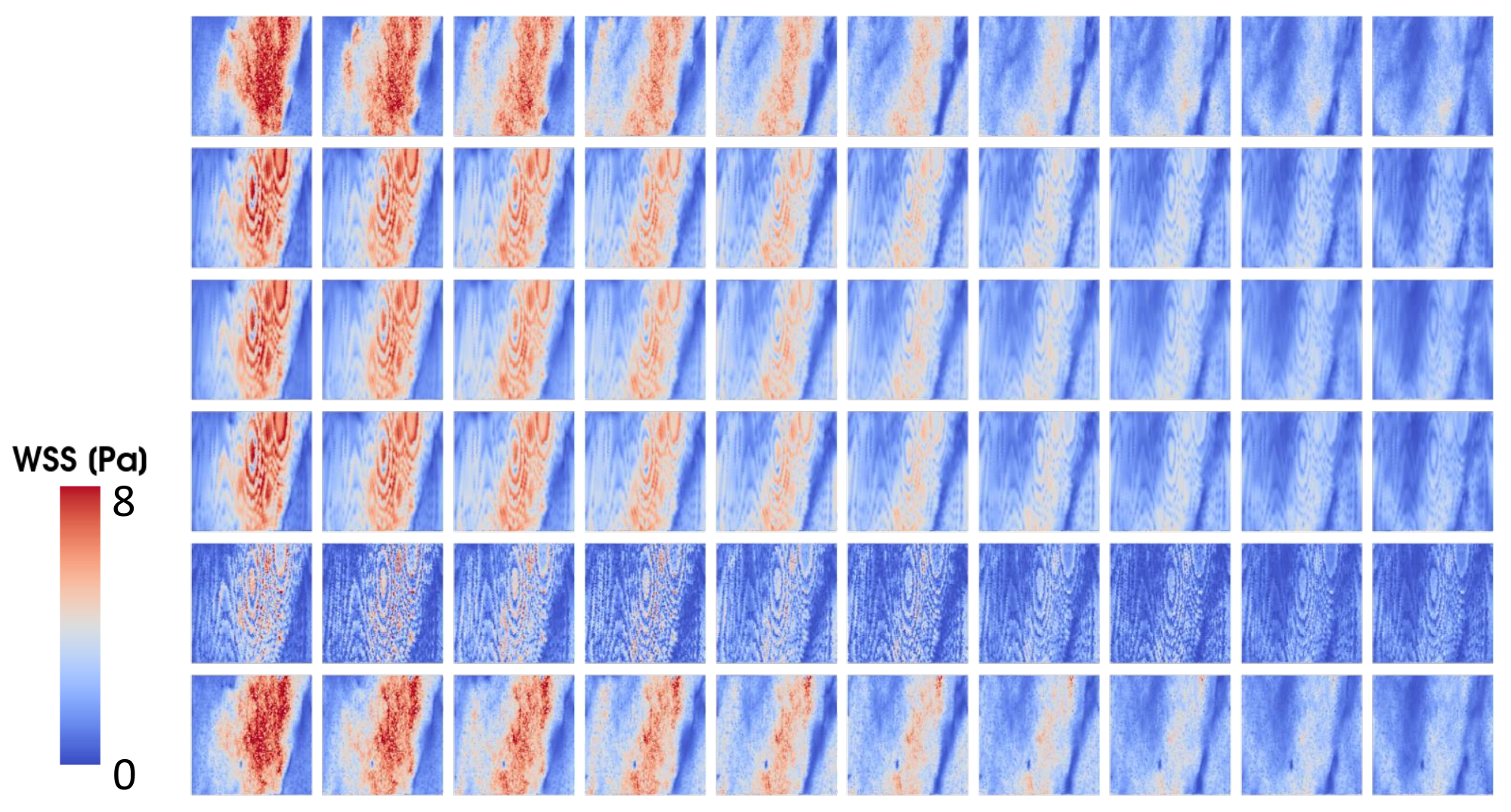}
    \caption{Results corresponding to \textit{mild} noise levels. Wall shear stress (WSS) colormaps on the aortic wall represented as a 2D surface at time points evenly spaced by $\Delta t = 20 ms$. Top row: WSS computed on ground truth velocity fields from CFD. LITP (second row), DFW (third row), DF-RBF (fourth row), 4D-RBF (fifth row), our method (sixth row).}
    \label{fig:compare_wss_mild_patch}
\end{figure}

\begin{figure}[htbp]
    \centering
    \includegraphics[width=\textwidth]{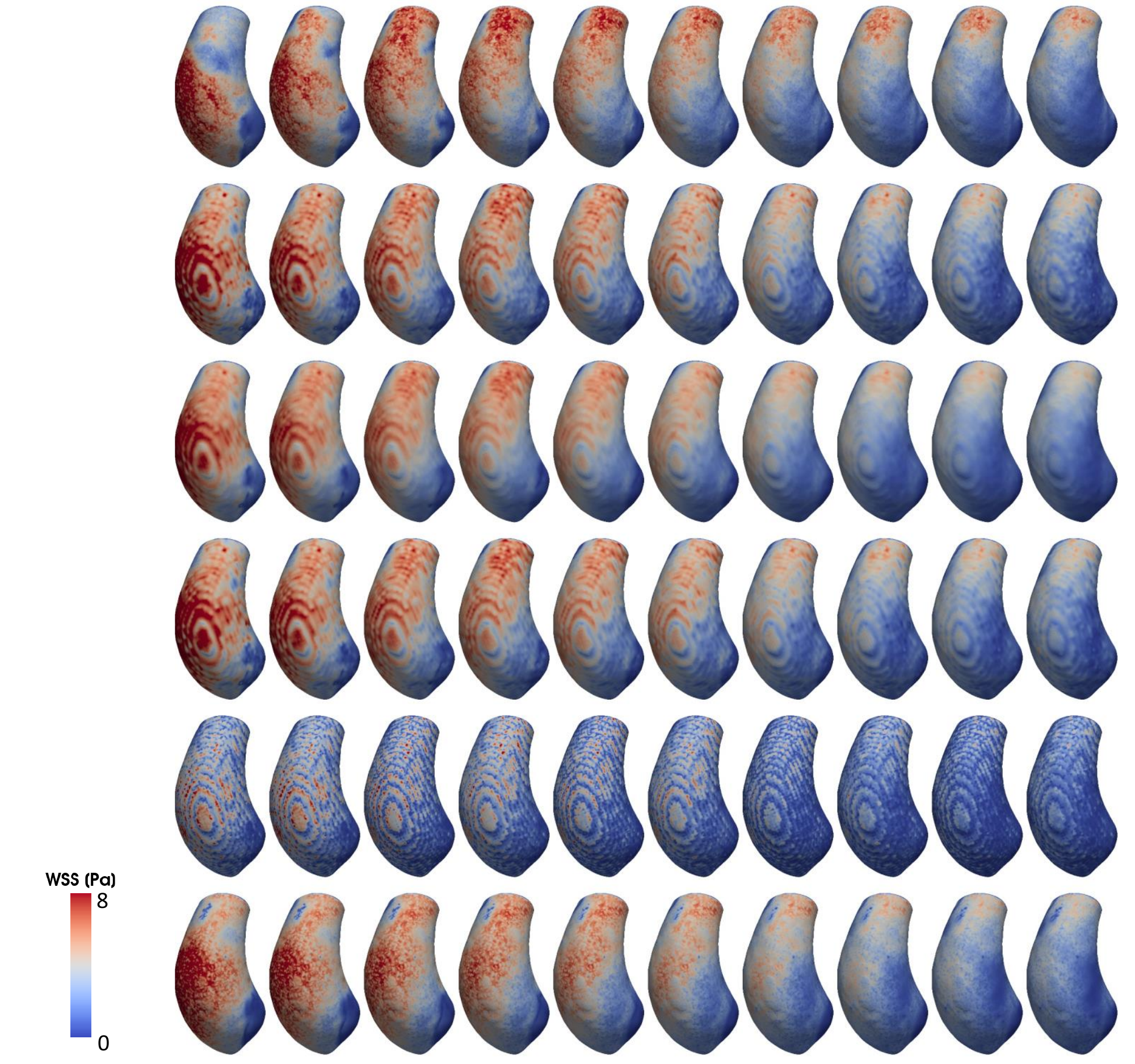}
    \caption{Results corresponding to \textit{medium} noise levels. Wall shear stress (WSS) colormaps on the aortic wall at time points evenly spaced by $\Delta t = 20 ms$. Top row: WSS computed on ground truth velocity fields from CFD. LITP (second row), DFW (third row), DF-RBF (fourth row), 4D-RBF (fifth row), our method (sixth row).}
    \label{fig:compare_wss_medium}
\end{figure}

\begin{figure}[htbp]
    \centering
    \includegraphics[width=\textwidth]{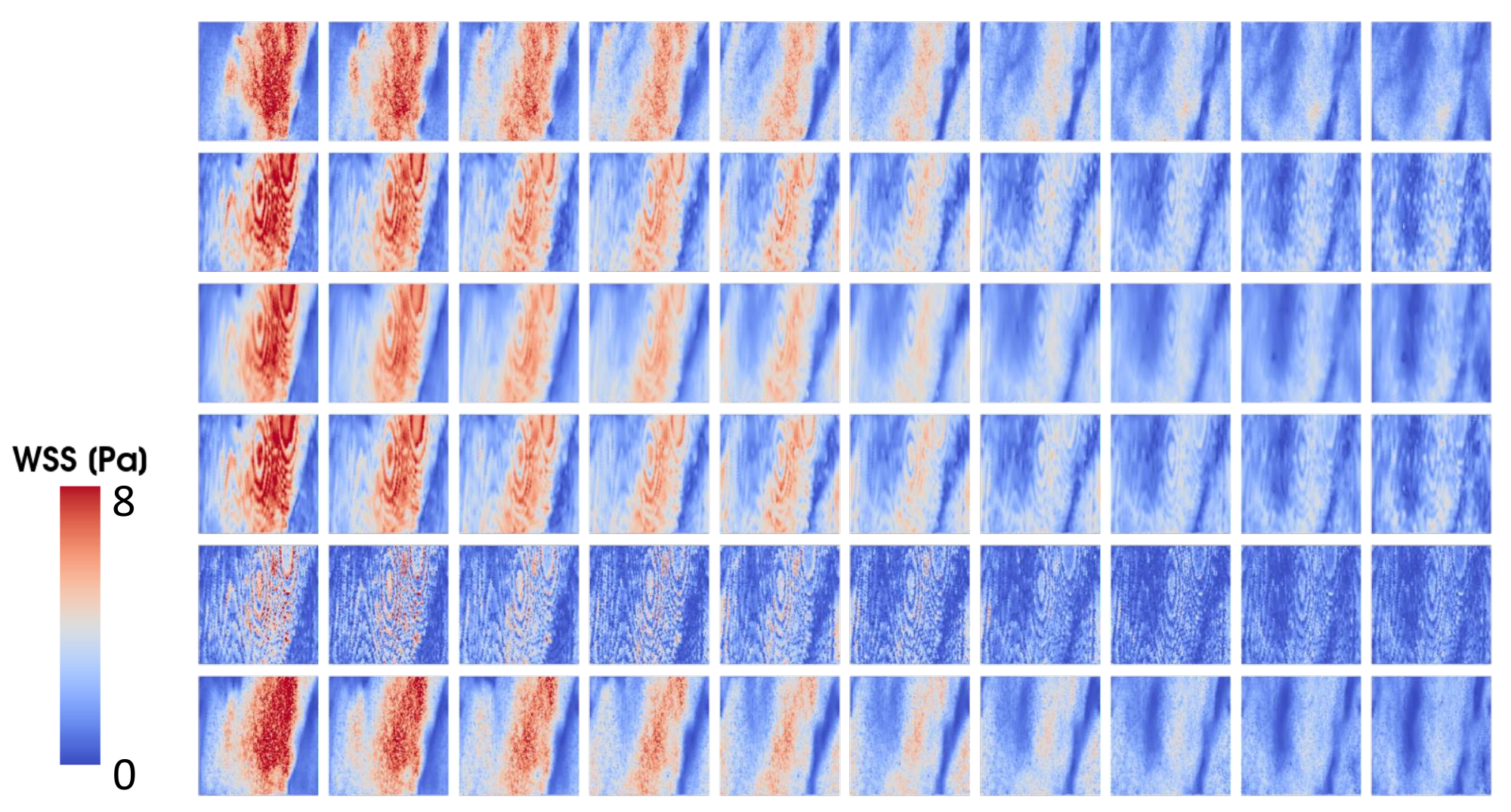}
    \caption{Results corresponding to \textit{medium} noise levels. Wall shear stress (WSS) colormaps on the aortic wall represented as a 2D surface at time points evenly spaced by $\Delta t = 20 ms$. Top row: WSS computed on ground truth velocity fields from CFD. LITP (second row), DFW (third row), DF-RBF (fourth row), 4D-RBF (fifth row), our method (sixth row).}
    \label{fig:compare_wss_medium_patch}
\end{figure}

\begin{figure}[htbp]
    \centering
    \includegraphics[width=\textwidth]{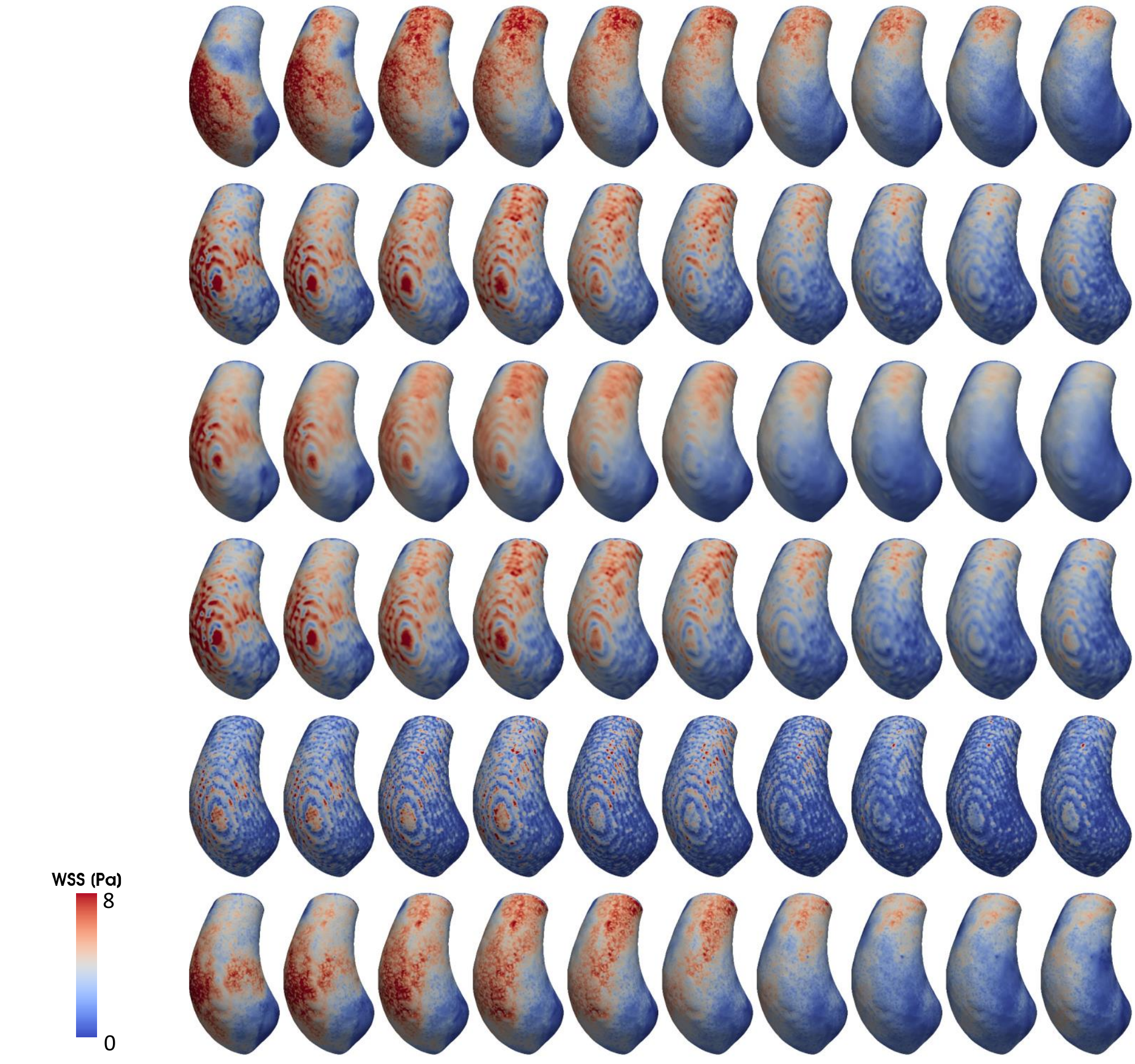}
    \caption{Results corresponding to \textit{extreme} noise levels. Wall shear stress (WSS) colormaps on the aortic wall at time points evenly spaced by $\Delta t = 20 ms$. Top row: WSS computed on ground truth velocity fields from CFD. LITP (second row), DFW (third row), DF-RBF (fourth row), 4D-RBF (fifth row), our method (sixth row).}
    \label{fig:compare_wss_extreme}
\end{figure}

\begin{figure}[htbp]
    \centering
    \includegraphics[width=\textwidth]{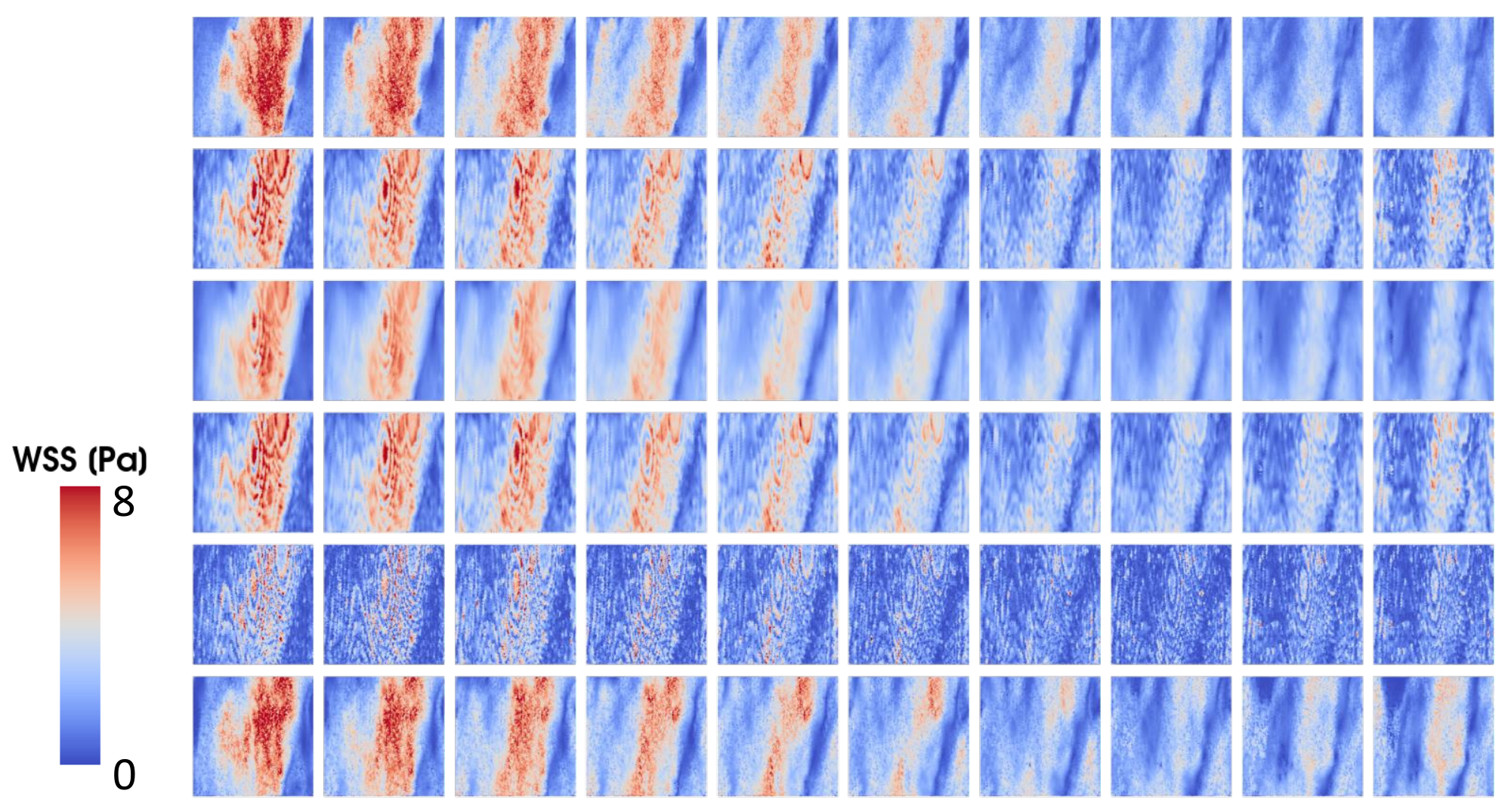}
    \caption{Results corresponding to \textit{extreme} noise levels. Wall shear stress (WSS) colormaps on the aortic wall represented as a 2D surface at time points evenly spaced by $\Delta t = 20 ms$. Top row: WSS computed on ground truth velocity fields from CFD. LITP (second row), DFW (third row), DF-RBF (fourth row), 4D-RBF (fifth row), our method (sixth row).}
    \label{fig:compare_wss_extreme_patch}
\end{figure}

The same methodology described in Section \ref{sec:wss_method} was applied to the velocity fields obtained by CFD (ground truth), DFW, DF-RBF, 4D-RBF and our best \textsc{siren} (20 layers, 300 neurons per layer). Results obtained by the different approach are reported in Table \ref{tab:comparison_results_wss}. Our method gave the lowest mNRMSE, vNRMSE and DE, outperforming the others for all the tested noise levels. WSS fields computed from the denoised and super-resolved velocity field obtained by the different approaches can be visualized as 3D colormaps in Figures \ref{fig:compare_wss_mild}, \ref{fig:compare_wss_medium} and \ref{fig:compare_wss_extreme} for \textit{mild}, \textit{medium} and \textit{extreme} noise levels, respectively, and as unwrapped 2D surfaces in figures \ref{fig:compare_wss_mild_patch}, \ref{fig:compare_wss_medium_patch} and \ref{fig:compare_wss_extreme_patch}.
More visible differences produced by the different methods can be appreciated for the \textit{extreme} noise level (Figures \ref{fig:compare_wss_extreme} and \ref{fig:compare_wss_extreme_patch}). In this case, all methods, including ours, underestimated WSS magnitudes. Our \textsc{siren} showed superior performance, giving WSS fields in good agreement with CFD-derived results, with high WSS magnitudes on the outer curve of the vessel wall, as expected.  

\begin{table}[htbp]
\centering
\begin{tabular}{cccccc}
\hline
 & LITP & \begin{tabular}[c]{@{}c@{}}DFW\\ +\\ LITP\end{tabular} & \begin{tabular}[c]{@{}c@{}}DF-RBF \\ +\\ LITP\end{tabular} & 4DRBF & \textsc{siren} \\ \hline
                    & 8.76 & 8.65 & 8.68 & 18.6 & 5.62  \\
\textit{mild}       & 6.03 & 6.10 & 6.00 & 12.31 & 4.02  \\
                    & 3.43 & 4.64 & 3.52 & 3.84 & 2.1  \\[0.2cm]
                    & 8.43 & 7.93 & 8.29 & 18.6 & 6.53  \\
\textit{medium}     & 6.05 & 5.99 & 5.96 & 12.6 & 4.76  \\
                    & 4.94 & 5.83 & 4.86 & 6.85 & 3.98  \\[0.2cm]
                    & 10.32 & 9.06 & 10.1 & 19.5 & 7.82  \\
\textit{extreme}    & 7.59 & 7.01 & 7.41 & 13.4 & 6.04  \\
                    & 7.74 & 8.41 & 7.5 & 10.34 & 7.0 \\ \hline
\end{tabular}
\caption{Wall shear stress field comparison with existing methods. In each cell values of mNRMSE (top), mNRMSE (middle) and DE (bottom) are reported.}
\label{tab:comparison_results_wss}
\end{table}

\FloatBarrier

\subsection{Case 2: velocity field assessment}
\begin{figure}[htbp]
    \centering
    \includegraphics[width=\textwidth]{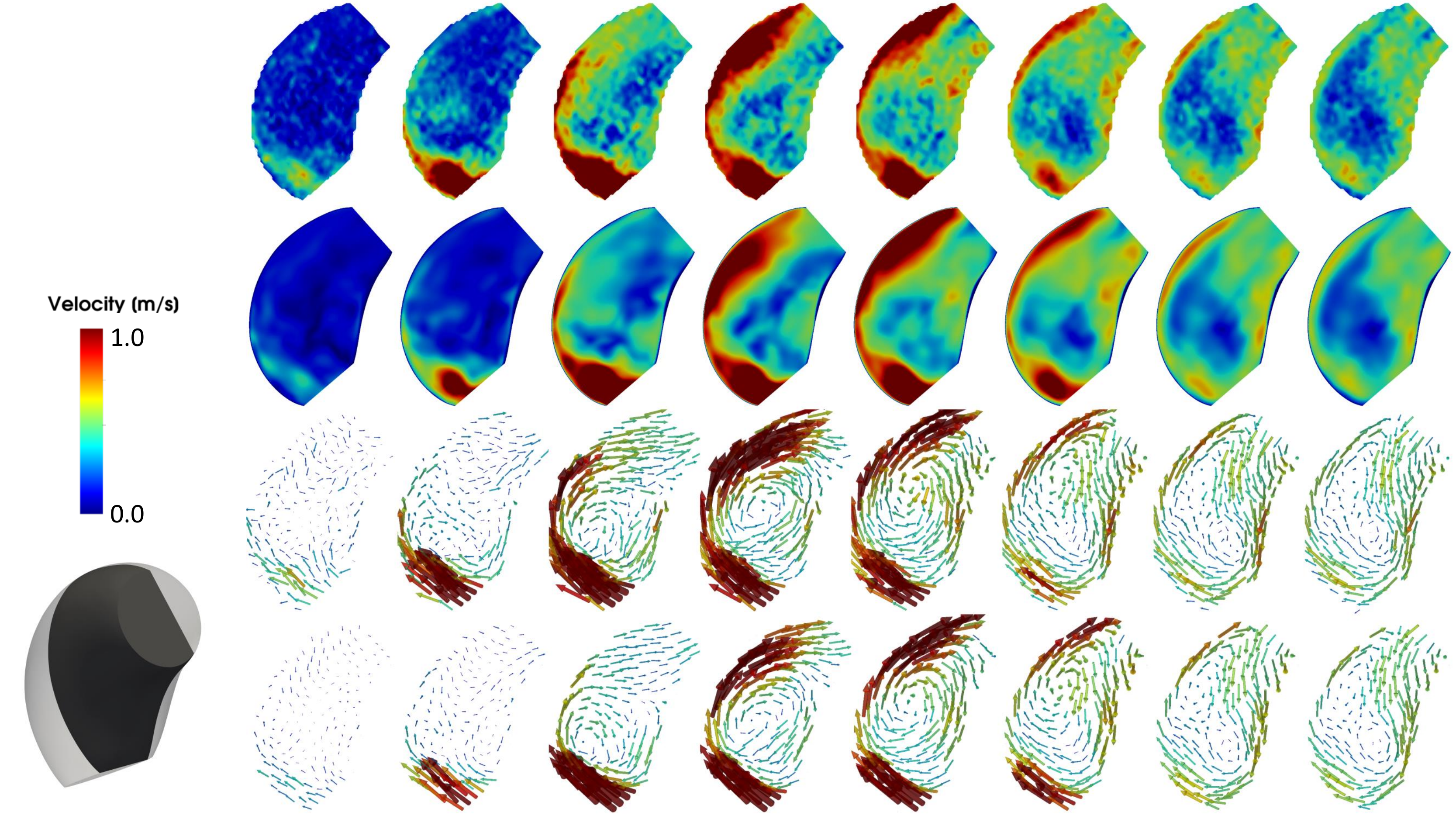}
    \caption{Velocity magnitude colormaps (rows 1 and 2) and velocity vectors (rows 3 and 4) for Case 2 on a sagittally oriented 2D slice (bottom left). Rows 1 and 3: 4D flow measurements, rows 2 and 4: \textsc{siren} results.}
    \label{fig:medical_vel}
\end{figure}

\begin{figure}[htbp]
    \centering
    \includegraphics[width=\textwidth]{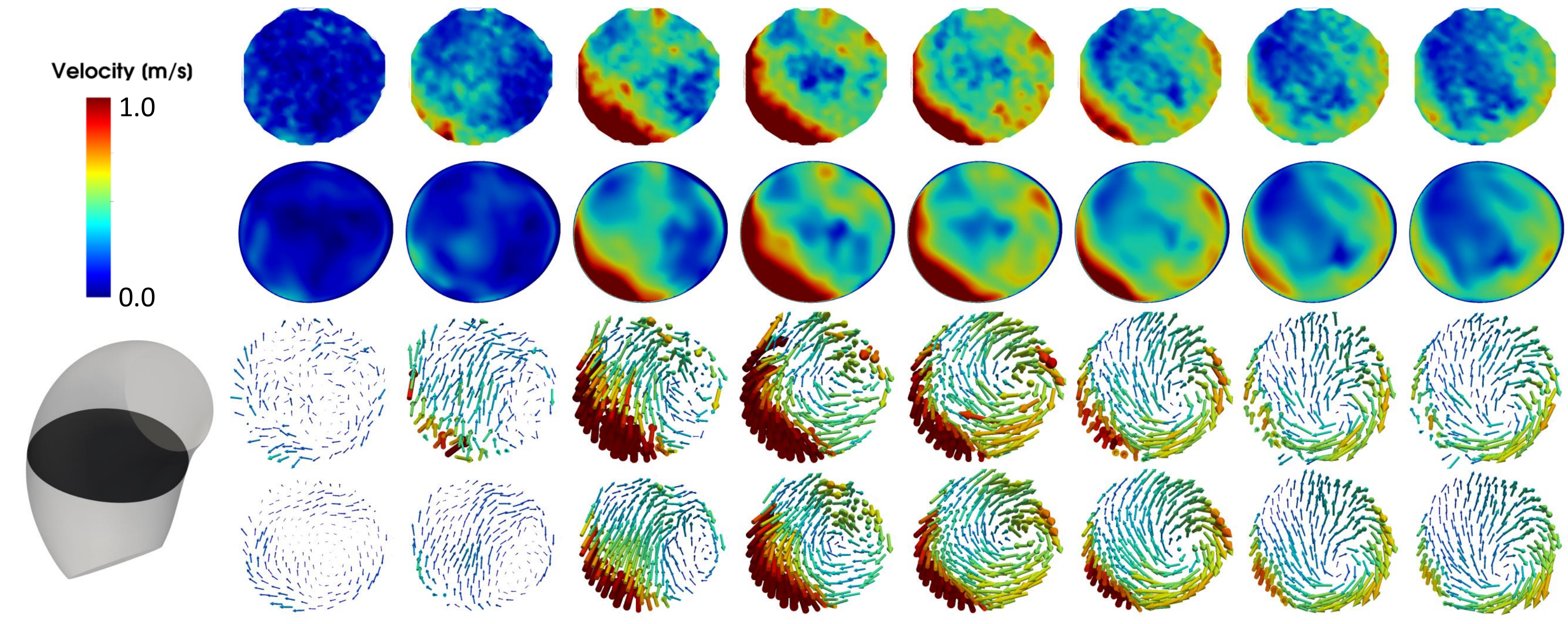}
    \caption{Velocity magnitude colormaps (rows 1 and 2) and velocity vectors (rows 3 and 4) for Case 2 on an axially oriented 2D slice (bottom left). Rows 1 and 3: 4D flow measurements, rows 2 and 4: \textsc{siren} results.}
    \label{fig:medical_vel_axial}
\end{figure}

Table \ref{tab:medical_measurements} shows quantitative measurements from raw MRI data and reconstructed velocity fields with the proposed \textsc{siren}. Overall, our method gave high resolution velocity fields that maintained low discrepancies in macroscopic quantitative measurements that are considered to be reasonably accurate when assessed from unprocessed MRI measurements \citep{ha2016hemodynamic}, giving differences in mean and maximum flow rate of \textless 5\% and underestimating reverse flow index (RFI) by 9.4\%, as computed in \citep{saitta2021qualitative}.
Qualitatively, our approach produced cleaner velocity fields, but maintaining the high velocity regions observed at the extrados of the ascending aorta as in the measured data (Figures \ref{fig:medical_vel} and \ref{fig:medical_vel_axial}). From the 3D vector visualization (third and fourth rows in Figures \ref{fig:medical_vel} and \ref{fig:medical_vel_axial}), it can be appreciated how the reconstructed velocity vector field kept the swirling patterns formed in the middle ascending aorta, but filtered out the spurious vector components in the near-wall regions.

\begin{table}[htbp]
\centering
\begin{tabular}{lccc}
\hline
 & \multicolumn{1}{c}{Measured data} & \multicolumn{1}{c}{INR} & \multicolumn{1}{c}{$\Delta \%$} \\ \hline
Mean flow rate {[}L/min{]} & 4.66 & 4.88 & 4.9\% \\
Max flow rate {[}L/min{]}  & 27.7 & 27.2 & 1.8\% \\
RFI {[}\%{]}               & 21 & 19 & 9.4\%\\ \hline
\end{tabular}
\caption{Quantitative measurements for Case 2 on raw flow MR measurements and on velocity fields reconstructed with our method.}
\label{tab:medical_measurements}
\end{table}

\subsection{Case 2: wall shear stress field assessment}
\begin{figure}[htbp]
    \centering
    \includegraphics[width=\textwidth]{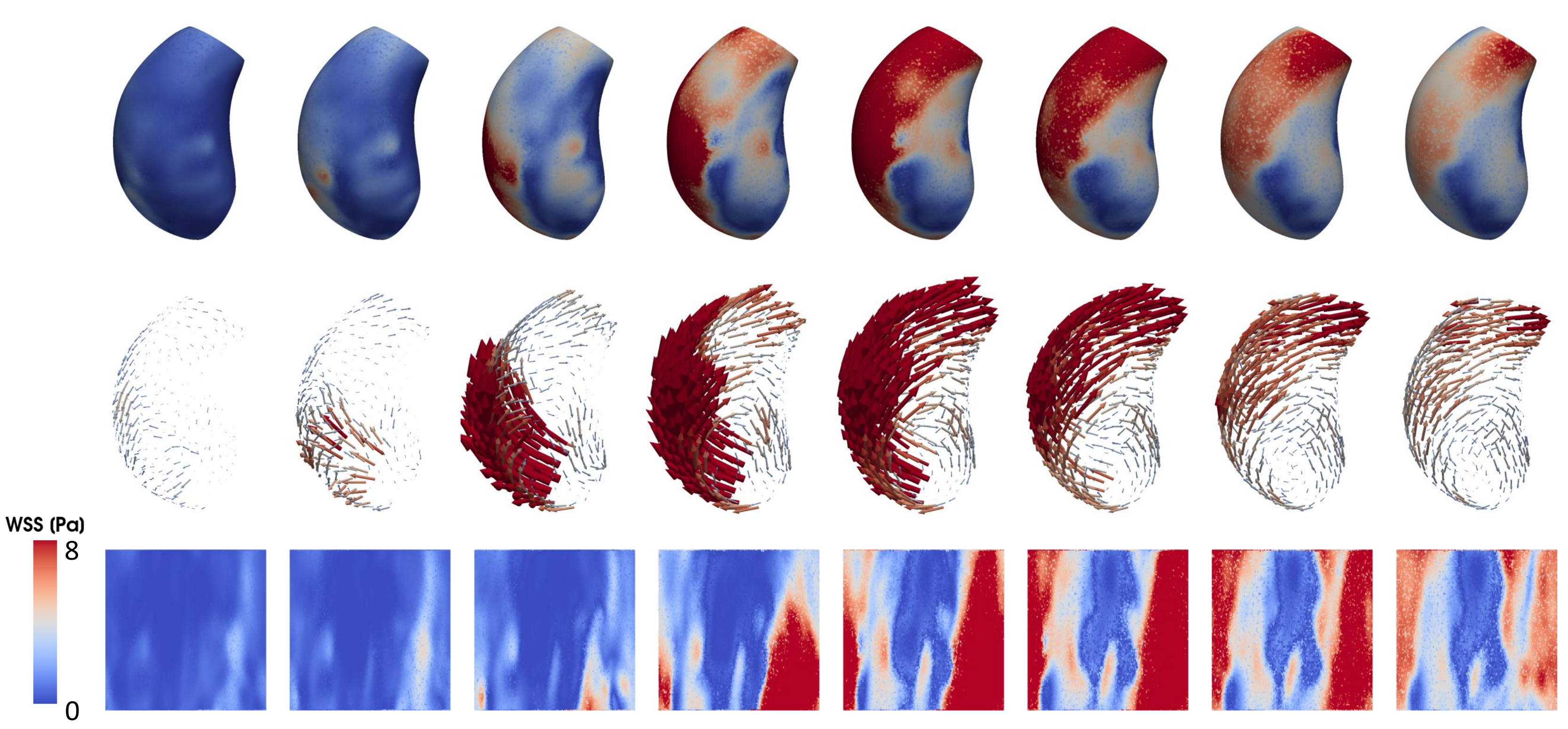}
    \caption{Wall shear stress (WSS) magnitude colormaps (row 1) and WSS vectors (rows 2) computed from \textsc{siren} results.}
    \label{fig:medical_wss}
\end{figure}

As observed for the synthetic case, the high velocity region near the wall of the outer curve of the aneurysm caused higher WSS in this region (Figure \ref{fig:medical_wss}), with a maximum WSS value of 20 Pa and mean time-averaged WSS (TAWSS) of 7.5 mPa. 

\FloatBarrier

\section{Discussion}
The use of 4D flow MRI in analyzing blood flow in major vessels has been widely studied, but the limitations of this imaging technique, such as noise and low spatio-temporal resolution have hindered its use in more advance velocity-based hemodynamic analysis. For instance, some studies have attempted to quantify WSS directly from flow-encoded MR images \citep{sotelo20183d, piatti20174d}, evaluating their accuracy at different levels of image noise or spatial resolution. Although these approaches enable fast assessment of near-wall quantities, they often tend to underestimate the true WSS values with respect to numerical simulation. In pursuit of more accurate quantification of WSS fields, researchers have turned to increasingly complex numerical simulations of blood flow \citep{funke2019variational, mut2011computational}. Although CFD studies are a powerful tool for estimating WSS, they require significant computational resources and can take several hours, or even days, to run due to their computational cost. \\
In this study, we proposed an unsupervised learning method based on INRs for denoising and SR of velocity fields measured by 4D flow MRI. We evaluated our approach on realistic synthetic data with various levels of noise and showed the superior performance of our method with respect to state-of-the-art methods in terms of both denoising and spatio-temporal SR. The proposed approach was able to denoise and super-resolve 4D velocity data while maintaining the integrity of the dominant flow features.

Among neural network approaches, it is worth mentioning the studies of Ferdian et al. \citep{ferdian20204dflownet}, Rutkowski et al. \citep{rutkowski2021enhancement} and \citep{shit2022srflow}. In these works, CFD simulations are used to create synthetic 4D flow MRI datasets, so to have measurements - ground truth pairs to train convolutional neural networks for denoising and SR. Hypothetically, if large realistic training data is generated, these approached would be able to learn the denoising and SR tasks, and could generalize to new domains without the need of re-training. Nonetheless, by operating convolutions in the image domain, these models are usually engineered to super-resolve $\times$2 or $\times$4 in space alone, and they do not allow for precise evaluation of near-wall quantities. Additionally, these methods require full ground truth supervision. For these reason, we did not include them in our comparison against existing methods.

The success of our method relies on two main properties of MLPs. First, we leverage \textsc{siren}'s spectral bias \citep{basri2020frequency, rahaman2019spectral} to achieve velocity field denoising. This property of dense fully-connected networks prevents them from learning high frequency functions. Assuming noise in MR to be Gaussian in \textit{k}-space results in high frequency artifacts in image space that are effectively removed by an INR. Our experiments on simulated 4D flow data (Case 1) allowed us to gain insights into the capabilities of \textsc{siren}s to denoise velocity data. When exploiting the spectral bias for signal denoising, one should carefully choose model's width and depth such that high frequency noise is filtered out, while desirable fine signal details are maintained. Overall, with \textit{mild} and \textit{medium} noise levels, all the tested combinations of depths and widths were able to fit the data, with a tendency of larger models to give slightly lower errors (Tables \ref{tab:hyperparams_results_mild}, \ref{tab:hyperparams_results_medium}). For \textit{extreme} noise levels, \textsc{siren}s with more than 300 neurons per layer overcame the spectral bias and overfitted high frequency noise. Once the best architecture was identified, our approach outperformed all other tested denoising methods.
The second key strength of the proposed approach relies on the fact that the trained MLP fits the velocity vector field as a continuous function of space-time coordinates. In practice, this was achieved by building upon the work of Sitzmann et al. \citep{sitzmann2020implicit}, who showed how \textsc{siren}s are better suited to fit complicated, feature-rich signals, such as natural images and solutions to simple PDEs. The present work is the first to adopt \textsc{siren}s for fitting 4D velocity measurements. By acting pointwise on 4D coordinates, a trained \textsc{siren} can be queried at continuous spatio-temporal locations, theoretically providing SR at arbitrarily fine spatial and temporal scales. Additionally, by incorporating time as an input feature, \textsc{siren}'s pointwise outputs are implicitly affected by temporal neighboring point features.

In principle, our approach is closely related to the formulation introduced by PINNs \citep{raissi2019physics}, with the main difference lying in the definition of the loss function. We train our models only using a data fidelity term, neglecting physical priors. The introduction of N-S residuals in differential form in the loss function could potentially lead to more physically consistent velocity fields, but would significantly slow down the training process. In the present study, we were interested in demonstrating the feasibility of a method that could potentially be applied to real medical scenarios where speed of execution is required. For the TAA patient (Case 2), using an NVIDIA A100 graphics card, our method took approximately 4 minutes to train and less than 2 seconds to evaluate at a fine spatio-temporal resolution.

In contrast to most data-based approaches, the devised method requires the definition of a bounded domain $\Omega \times [t_a, t_b]$ (Section \ref{sec:training}). On one hand, this choice represents a limitation of our workflow, entailing longer processing times. On the other hand, precise definition of a smooth vessel wall surface enables robust computation of WSS, a clinically important hemodynamic biomarker. Results on synthetic data showed good quantitative and qualitative agreement between predicted and reference CFD data (Table \ref{tab:comparison_results_wss} and Figures \ref{fig:compare_wss_mild}, \ref{fig:compare_wss_medium}, \ref{fig:compare_wss_extreme}). Results on real medical data revealed a high WSS region on the outer curvature of the ascending aorta (Figure \ref{fig:medical_wss}), with most values ranging from 0 to 15 Pa. These results are in good agreement with a recent study on TAA biomechanics \citep{salmasi2021high}, who reported maximum WSS values of 10.18 $\pm$ 4.14 Pa for a cohort of 10 patients. Even though more detailed analyses are needed on the use of neural networks for \textit{in vivo} WSS estimation, our results indicate that velocity fields produced by our method are suitable for extraction of derived biomarkers, otherwise difficult to assess.

\section{Conclusions}
In this work we showed the feasibility of \textsc{siren}s to represent complex, high dimensional blood flow velocity fields measured by 4D flow MRI. By training on low resolution coordinates, our method is quick to execute for new cases and easy to implement. By carefully tuning our \textsc{siren} architecture, we exploit the spectral bias to obtain a functional representation of our data with reduced noise, outperforming state-of-the art solutions. Our method provides continuous velocity fields that can be queried at arbitrary spatio-temporal locations, effectively achieving 4D super-resolution.

\clearpage
\bibliographystyle{ieeetr}
\bibliography{references} 
\end{document}